\documentclass[pdftex]{pasj01}

\usepackage{graphics,times}
\usepackage{url}

\usepackage{ulem}

\usepackage{lineno}
\usepackage{natbib}

\begin{document} 
\Received{}
\Accepted{}

\title{Exploring the cosmic dawn and epoch of reionization with 21cm line}


\author{Hayato Shimabukuro\altaffilmark{1}%
\thanks{shimabukuro@ynu.edu.cn}}
\author{Kenji \textsc{Hasegawa}\altaffilmark{2}}%
\author{Akira \textsc{Kuchinomachi}\altaffilmark{3}}%
\author{Hidenobu \textsc{Yajima}\altaffilmark{4}}%
\author{Shintaro \textsc{Yoshiura}\altaffilmark{5}\altaffilmark{6}}%
\altaffiltext{1}{Yunnan university,SWIFAR,No. 2 North Green Lake Road, Kunming, Yunnan Province 650500, China}
\altaffiltext{2}{Department of Physics and Astrophysics, Nagoya University Furo-cho, Chikusa-ku, Nagoya, Aichi 464-8602, Japan}
\altaffiltext{3}{Kumamoto University, Faculty of Advanced Science and Technology, Kumamoto, Kumamoto 860-8555, Japan}
\altaffiltext{4}{Center for Computational Sciences, University of Tsukuba, Ten-nodai, 1-1-1 Tsukuba, Ibaraki 305-8577, Japan}
\altaffiltext{5}{Mizusawa VLBI Observatory, National Astronomical Observatory Japan, 2-21-1 Osawa, Mitaka, Tokyo 181-8588, Japan}
\altaffiltext{6}{The University of Melbourne, School of Physics, Parkville, VIC 3010, Australia}



\KeyWords{dark ages, reionization, first stars} 

\maketitle

\begin{abstract}

The dark age of the universe, when no luminous object had existed, ended with the birth of the first stars, galaxies, and blackholes. This epoch is called cosmic dawn. Cosmic reionization is the major transition of the intergalactic medium (IGM) in the universe driven by ionizing photons emitted from luminous objects. Although the epoch through the dark age to reionization is a milestone in the universe, our knowledge of this epoch has not been sufficient yet. Cosmic 21cm signal, which is emitted from neutral hydrogen, is expected to open a new window for this epoch. In this review paper, we first introduce the basic physics of the 21cm line and how first stars impact on the 21cm line signal. Next, we briefly summarize how we extract astrophysical information from the 21cm line signal by means of statistical and machine learning approaches. We also discuss the synergy between the 21cm line signal and other emission lines. Finally, we summarize the current status of 21cm experiments.

\end{abstract}


\section{Introduction}
At the beginning of the universe, there had been no luminous objects such as stars and galaxies. We call this epoch “Dark ages”. Based on the hierarchical structure formation scenario of the standard cosmology model, dark matter mini-halos had formed, and the first stars(Pop.III stars) formed in the mini-halos\citep[e.g.][]{2003ApJ...592..645Y,2006ApJ...652....6Y,2008Sci...321..669Y}. The first generations of stars, galaxies, and black holes are expected to have formed by earlier than 1 billion year after the Big bang \citep{2001PhR...349..125B}. This epoch is called ``cosmic dawn”. Theoretical studies indicate that high energetic photons emitted from such black holes or X-ray binaries at the very early Universe heated the Universe. The bulk of hydrogen atoms had kept a neutral state before the first stars and galaxies formed. However, the phase of the Universe had drastically changed by ultraviolet (UV) photons emitted from the first stars in the first galaxies. These UV photons ionized neutral hydrogen atoms in the Intergalactic medium (IGM) and then the whole of the Universe had been gradually ionized\citep[e.g.][]{2006ARA&A..44..415F}. This epoch is called ``Epoch of Reionization (EoR)''. The EoR is one of the mysterious epochs in the Universe.  

Planck satellites measured the optical depth of Thomson scattering of Cosmic Microwave Background (CMB) photons,$\tau_{e}=0.066 \pm 0.016$, which corresponds to instantaneous reionization redshift $z_{r}=8.8^{+1.7}_{-1.4}$ \citep{2016A&A...594A..13P}. 

For the last 20 years, exciting optical/UV telescopes, such as the Subaru telescope and Hubble telescope, have detected high redshift galaxies at $z > 6$ \citep[see review papers, e.g.][]{2006ARA&A..44..415F,2020ARA&A..58..617O}. We observe a steep faint-end slope of the UV luminosity function at $z \sim 6 - 8$ down to an absolute UV magnitude of $M_{\mathrm{UV}}\sim -16$ \citep{2015ApJ...800...18A,2015ApJ...799...12I}. The observations of UV luminosity function give clues about the property of ionizing source such as ionizing efficiency \citep[e.g.][]{2015ApJ...802L..19R,2015ApJ...811..140B,2018ApJ...854...73I}. These results imply that high-redshift galaxies, in particular, faint ones are the main candidates for ionizing photon sources which occur in neutral hydrogen reionization. Lyman alpha emitter (LAE) galaxies have been one of the major observational probes of high redshift Universe. Since the Lyman alpha line is emitted around the star formation region, LAE is considered as young and high star formation rate galaxies. The observations of LAE galaxies luminosity function put constraints on the neutral fraction of IGM during the late stage of reionization around $z\sim 6-7$ \citep[e.g.][]{2010ApJ...723..869O, 2014ApJ...797...16K,2018PASJ...70S..16K}. Absorption signatures in the spectra of quasars are also one of the probes to investigate the nature of IGM. Especially, a series of absorption lines seen at a wavelength shorter than the Lyman alpha line (in the rest frame) is called the Lyman alpha forest. The Lyman alpha forest seen in quasar spectrum at $z>7$ has constrained the ionized history of the IGM at the late stage of EoR \citep[e.g.][]{2018Natur.553..473B,2019MNRAS.484.5094G,2019ApJ...878...12H,2020ApJ...897L..14Y}.  A zero-flux pixel in the Lyman alpha or Lyman beta forests appeared in high-$z$ quasar spectrum, is called dark pixel\citep{2010MNRAS.407.1328M}. The dark pixel results from either a fully neutral region or residual neutral region inside the ionized IGM. The dark fraction(fraction of the dark pixels) provides the (nearly) model-independent upper limit on the neutral hydrogen fraction \citep[e.g.][]{2011MNRAS.415.3237M,2015MNRAS.447..499M}. Recently, extremely long($\sim 110 h^{-1}{\rm Mpc}$) and dark (the effective optical depth $\tau_{{\rm eff}}$ is larger than 7) Lyman alpha trough at $z\sim 5.5-6.0$ has been reported \citep{2015MNRAS.447.3402B,2018ApJ...863...92B}. To explain this extreme Lyman alpha trough, some scenarios, such as fluctuating ultraviolet background driven by galaxies model\citep[e.g.][]{2016MNRAS.460.1328D,2018MNRAS.473..560D}, fluctuating temperature model\citep[e.g.][]{2015ApJ...813L..38D} and ultra-late reionization model\citep[e.g.][]{2019MNRAS.485L..24K,2020MNRAS.491.1736K}, are suggested. However, the origin of the trough has been under debate.

Apart from optical/UV wavelength, radio observations are also powerful for studying high redshift galaxies. Atacama Large Millimeter Array (ALMA) has been making outstanding progress in observing millimeter/sub-millimeter galaxies. Some of ALMA observations reported the detection of dust continuum in galaxies at $z>7$ \citep[e.g.][]{2015Natur.519..327W,2017ApJ...837L..21L,2019ApJ...874...27T}. Since dust grains are mainly formed by condensation of heavy elements, the observation of dust gives a clue of the metallicity environment in the galaxies during the EoR. In addition to dust continuum, ALMA has also detected [CI\hspace{-.1em}I] 158$\mu $m and [OI\hspace{-.1em}I\hspace{-.1em}I] 88$\mu $m  emission lines in high redshift galaxies at $z > 7$ \citep[e.g.][]{2016ApJ...829L..11P,2016Sci...352.1559I,2018Natur.557..392H,2019PASJ...71...71H,2020ApJ...896...93H}. In particular, \citet{2019PASJ...71...71H} reported the detection of [OI\hspace{-.1em}I\hspace{-.1em}I], [CI\hspace{-.1em}I] emission lines and dust continuum from the galaxy at $z=7.15$. Since [CI\hspace{-.1em}I] is one of the main Interstellar medium (ISM) cooling lines and the brightest far-infrared line in star-forming galaxies, [CI\hspace{-.1em}I] is considered as a tracer of the star formation rate (SFR) of the galaxies and their gas dynamics. While [CI\hspace{-.1em}I] emission line mainly arises from neutral ISM/photo-dissociation regions, [OI\hspace{-.1em}I\hspace{-.1em}I] emission line arises from HII region and it is bright in the young galaxy. The ratio of  [OI\hspace{-.1em}I\hspace{-.1em}I]/[CI\hspace{-.1em}I]  provides us invaluable information on the chemical and ionization properties of galaxies. For example, low metallicity galaxies and highly ionized galaxies trend to be high [OI\hspace{-.1em}I\hspace{-.1em}I]/[CI\hspace{-.1em}I] ratio\citep[e.g. see reference in][]{2019PASJ...71...71H}. 

Both the Lyman alpha emission line and the [CI\hspace{-.1em}I] emission line are tracers of star-forming galaxies, but recently, by combining Subaru and ALMA data, the results of anti-correlation between [CI\hspace{-.1em}I]/SFR and Lyman alpha equivalent width have been obtained \citep{2018ApJ...859...84H,2020ApJ...896...93H}. This result is likely a consequence of high ionization parameter with strong radiation at high redshift galaxy and/or [CI\hspace{-.1em}I] emission coming from high-density photo-dissociation regions in the galaxies. However, it is still under debate.

As shown above, the observations of high redshift galaxies not only confirm sources at the EoR but also provide physical properties of the galaxies at the EoR, and also give constraints on the ionization history of the IGM. However, it is difficult to obtain detailed information on the structure of the ionized region in the IGM from such observations, and other observations are needed to learn the IGM from the dark ages to the EoR. 

A promising tool to investigate the IGM from the dark ages to the EoR more directly is the cosmic 21cm line signal, which is a spectral line emitted from neutral hydrogen atoms \citep{1990MNRAS.247..510S,1997ApJ...475..429M}. Since the dominant baryonic matter in the Universe is neutral hydrogen atoms, we can easily trace the ionized (neutral) structure of the IGM along with redshift via a 21~cm line signal. To detect the 21cm line signal, some radio interferometric telescopes such as MWA\citep[e.g.][]{2018PASA...35...33W}, LOFAR\citep[e.g.][]{2013A&A...556A...2V} and HERA\citep[e.g.][]{2017PASP..129d5001D} have already started observation. Recently, HERA gives tighter upper limits on the 21cm line power spectrum and constrains the cosmological and astrophysical models obtained by the HERA observation \citep{2021arXiv210807282T}. 

Independent of the interferometric observations, single dish type radio telescopes such as the Experiment to Detect the Global Epoch of Reionization Signature (EDGES) \citep{2018Natur.555...67B}, the Large-aperture Experiment to Detect the Dark Ages (LEDA)\citep{2018MNRAS.478.4193P}, the Probing Radio Intensity at high $z$ from Marion(PRIZM)\citep{2019JAI.....850004P} and the Shaped Antenna measurement of the background RAdio Spectrum2 $\&$ 3 (SARAS2$\&$3)\citep{2018ApJ...858...54S,2021arXiv210401756N} are targeting to detect the 21cm line global signal, which is sky-averaged 21cm line signal. Recently, \citet{2018Natur.555...67B} has reported the detection of 21cm line absorption line feature at 78 MHz. However, the reported absorption trough is too deep to be explained by standard cosmology and astrophysics scenarios \citep[e.g.][]{2017MNRAS.472.1915C}. To explain the absorption trough, many scenarios beyond standard cosmology and astrophysics, such as alternative dark matter scenario\citep[e.g.][]{2018Natur.555...71B,2018PhRvL.121a1101F,2018Natur.557..684M} and excess radio background model\citep[e.g.][]{2019MNRAS.486.1763F,2020MNRAS.499.5993R}, have been suggested. The excess radio background predicts additional fluctuations in the 21cm fluctuations\citep{2020MNRAS.499.5993R}. On the other hand, there is the argument that the EDGES results do not require a model beyond the standard cosmology and astrophysics, but that there was an oversight in the analysis of systematic errors\citep[e.g.][]{2019ApJ...874..153B,2019ApJ...880...26S,2020MNRAS.492...22S}. Further, SARAS3 has reported no detection of the 21cm global signal\citep{2022NatAs.tmp...47S}. Thus, the detection of the 21cm global signal reported by EDGES is still under debate.

If we detect the 21cm line signal, it enables us to explore the physics of the EoR and cosmic dawn ({\it e.g.} star formation rates, ionizing escape fraction, and galaxy X-ray property) \citep[e.g.][]{2015MNRAS.449.4246G,2019MNRAS.484..933P}. Furthermore, in the next decade, Square Kilometre Array (SKA) is expected to revolutionize our understanding of the epoch through the dark ages to the EoR with its unprecedented sensitivity\citep[e.g.][]{2015aska.confE...1K}

In this paper, we review the current progress of EoR and 21cm line studies. In section 2, we review the basic physics of the 21cm line. Readers can learn what physical mechanism determines the 21cm line signal. As astrophysics which affects the 21cm line signal, we focus on the impact of first stars on the 21cm line signal in section 3. In section 4, we review how we extract astrophysical information from the 21cm line signal statistically. Not only the 21cm line signal, but other lines are also useful to extract information during EoR. In section 5, we introduce synergy between the 21cm line signal and other lines. Finally, we review the current 21cm line experiments status in section 6.

\section{Cosmic 21cm signal}
The bulk of IGM before the EoR consists of the neutral hydrogen atom. The 21cm signal is emitted from a neutral hydrogen atom due to the hyperfine structure. The 21cm line signal is a powerful probe to explore the epochs through the dark ages to EoR. In this section, we summarize the basis of the 21cm line signal.

\subsection{Basic physics of the 21cm line}
Here, we introduce basic physics of the 21cm line. Based on \citet{1986rpa..book.....R}, we start from radiative transfer equation to describe the propagation of the radiation in the IGM.  The radiative transfer equation for an infinitesimal distance $ds$ is written by 
\begin{equation}
\frac{dI_{\nu}}{ds}=-\alpha_{\nu}I_{\nu}+j_{\nu},
\label{eq:radiative}
\end{equation}
where subscript $\nu$ denotes the frequency.  $I_{\nu}$ is specific intensity of incident light and $\alpha_{\nu}$ is absorption coefficient. The incident light passing through IGM is absorbed and intensity of incident is decreased by $\alpha_{\nu}I_{\nu}$.  $j_{\nu}$ is the emission coefficient per volume per solid angle.  Here, we define new quantities, optical depth $\tau$ and source function $S_{\nu}$. The (infinitesimal) optical depth $d \tau_{\nu}$ is defined by $d\tau_{\nu}\equiv\alpha_{\nu}{\it ds}$ and $S_{\nu}$ is defined by $S_{\nu}\equiv j_{\nu}/\alpha_{\nu}$. The source function $S_{\nu}$ is expressed by Planck function $B_{\nu}$ in thermal equilibrium state.  With these quantities, eq.(\ref{eq:radiative}) is re-written by 

\begin{equation}
\frac{d I_{\nu}}{d\tau_{\nu}}=-I_{\nu}+B_{\nu}
\label{eq:radiative2}
\end{equation}

\begin{figure}
    \includegraphics[width=1.0\hsize]{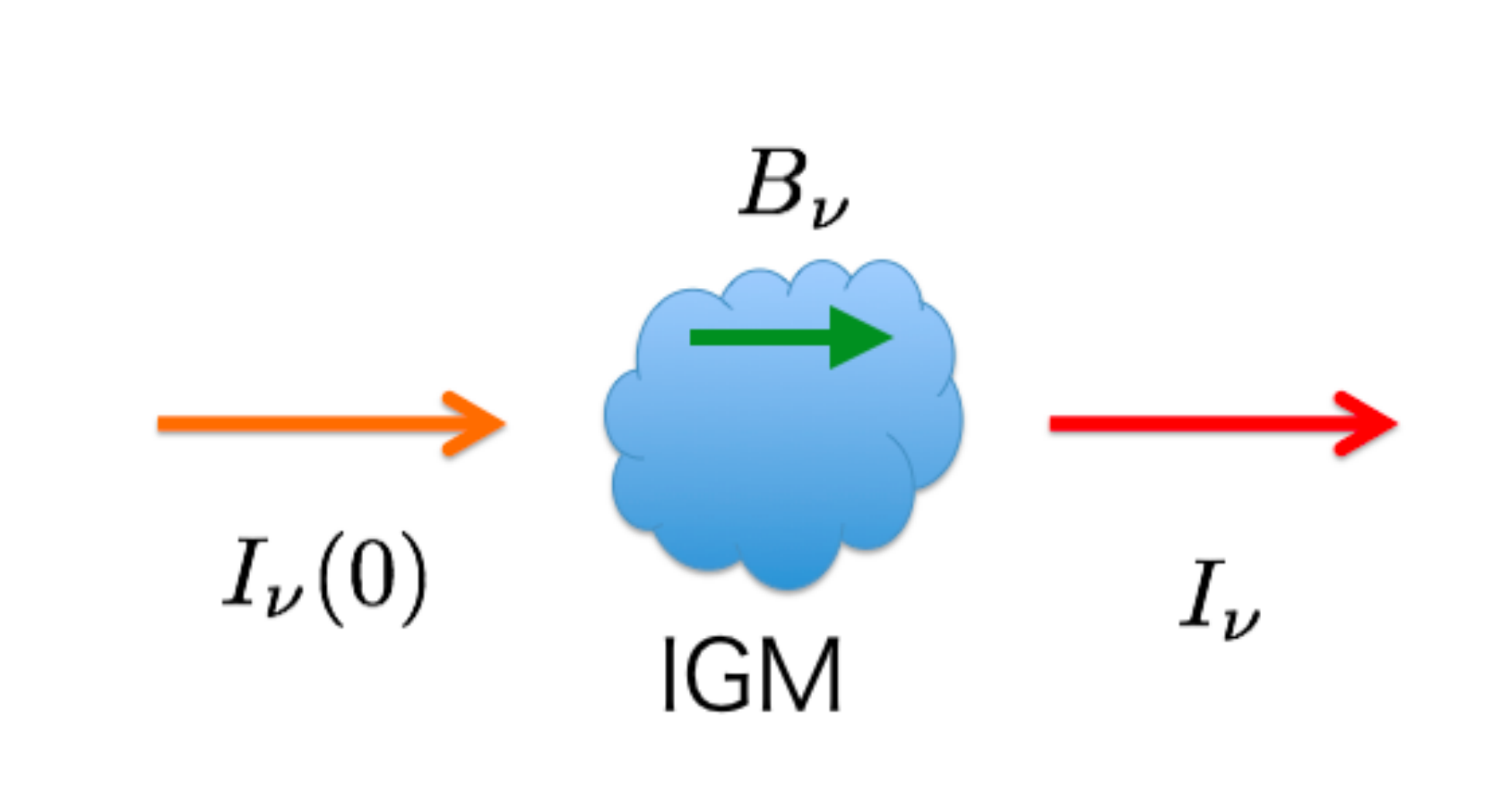}
    \caption{A schematic figure of the radiative transfer}
    \label{fig:radiation}
\end{figure}

Eq.\ref{eq:radiative2} gives a solution as follow

\begin{equation}
I_{\nu}=I_{\nu}(0)\exp(-\tau_{\nu})+B_{\nu}[1-\exp(-\tau_{\nu})].
\label{eq:radiative3}
\end{equation}
The physical meaning of this equation is that first term expresses the extinct incident absorbed by IGM with optical depth $\tau_{\nu}$ and second term shows extinct emission from source. We show schematic figure in Fig.\ref{fig:radiation}.

In the condition of thermal equilibrium state and low frequency regime relevant to 21cm line, we can use Rayleigh-Jeans approximation. Thus, $I_{\nu}$, $I_{\nu}(0)$ and $B_{\nu}$ become  $I_{\nu}=2k_{{\rm B}}T_{{\rm b}}\nu^{2}/c^{2}$, $I_{\nu}(0)=2k_{{\rm B}}T_{R}(\nu)\nu^{2}/c^{2}$, $B_{\nu}=2k_{{\rm B}}T_{{\rm ex}}\nu^{2}/c^{2}$, respectively.  $T_{b}$, $T_{{\rm ex}}$ and  $T_{R}$ are called brightness temperature, excitation temperature and brightness of radio background source, respectively. With these temperatures, eq.\ref{eq:radiative3} can be re-written by
\begin{equation}
T_{{\rm b}}=T_{R}(\nu)\exp(-\tau_{\nu})+T_{{\rm ex}}[1-\exp(-\tau_{\nu})]
\label{eq:radiative4}
\end{equation}
In the case of hypefine structure, excitation temperature $T_{{\rm ex}}$ is replaced with spin temperature $T_{{\rm S}}$. In the cosmological context, the brightness temperature of background is CMB temperature $T_{\gamma}$. The spin temperature $T_\mathrm{S}$ is defined by the ratio between the number density $n_{i}$ of hydrogen atom in the two hyperfine levels (singlet $n_{0}$ and triplet $n_{1}$),

\begin{equation}
    \frac{n_{1}}{n_{0}}=\frac{g_{1}}{g_{0}}\exp \left(\frac{-h\nu_{21}}{k_\mathrm{B} T_\mathrm{S}}\right)
\end{equation}
where, $(g_{1}/g_{0})=3$ is the ratio of statistical degeneracy of the singlet and triplet, $k_{\mathrm{B}}$ is Boltzmann constant and $h$ is Planck constant. $\nu_{21}= 1.42$GHz is the frequency corresponding to 21cm wavelength.

We observe the contrast between hydrogen clouds and the CMB. Thus, we measure differential 21cm brightness temperature \citep[e.g.][]{1990MNRAS.247..510S,1997ApJ...475..429M,Furlanetto:2006jb,2013ExA....36..235M}. The 21cm brightness temperature is given by

\begin{eqnarray}
\delta T_{b}(\mathbf{x},z) &=& \frac{T_{\mathrm{S}}-T_{\gamma}}{1+z}\tau_{\nu}\\
&=&27x_{{\rm HI}}(\mathbf{x},z)[1+\delta_{m}(\mathbf{x},z)] \nonumber\\
&\times& \bigg(1-\frac{T_{\gamma}(z)}{T_{{\rm S}}({\mathbf{x}},z)}\bigg)\left(1+\frac{1}{H(z)}\frac{dv_{||}}{dr_{||}}\right)^{-1} \nonumber\\
&\times&\bigg(\frac{1+z}{10}\frac{0.15}{\Omega_{m}h^{2}}\bigg)^{\frac{1}{2}}\bigg(\frac{\Omega_{b}h^{2}}{0.023}\bigg) [\mathrm{mK}],
\label{eq:brightness}
\end{eqnarray}
where $T_{\rm S}$ is the spin temperature of the IGM, $T_{\gamma}$ is the CMB temperature, ${dv_{||}}/{dr_{||}}$ is a peculiar velocity along line of sight.
$x_{\rm HI}(\mathbf{x},z)$ is neutral fraction of the hydrogen atom gas, and $\delta_{m}(\mathbf{x},z)$ is matter density fluctuation. Others are cosmological parameters.  All variables are evaluated at the redshift $z = \nu_{21}(=1.42{\rm GHz})/\nu - 1$. 

\subsection{Spin temperature}
In this subsection, we describe how we determine the spin temperature\citep[e.g.][]{2006PhR...433..181F,2012RPPh...75h6901P}. The spin temperature is determined by (1) absorption of the CMB photons by neutral hydrogen atom, (2) collisions with other hydrogen atoms and free electrons (3) resonant scattering of Lyman-$\alpha$ photons. Here, we let $C_{10}$ and $P_{10}$ be de-excitation rates by collisions and UV photon scattering, respectively.  We also let $C_{01}$ and $P_{01}$ be excitation rates.
In the equilibrium state the spin temperature is  determined by the balance between excitation and de-excitation described by
\begin{equation}
n_{1}(C_{10}+ P_{10}+A_{10}+B_{10}I_{{\rm CMB}})=n_{0}(C_{01}+P_{01}+B_{01}I_{{\rm CMB}}),
\label{eq:equilibrium}
\end{equation}
where $A_{10}, B_{01}, B_{10}$ are Einstein coefficients.  $A_{10}$ coefficient denotes the spontaneous emission and $B_{01}, B_{10}$ denotes photon absorption and induced emission, respectively.  $I_{{\rm CMB}}$ is the intensity of CMB photon.  The first and second terms on the left-hand side express the transition from triplet to singlet due to the collision and UV scattering respectively, third and fourth terms express the transition from triplet to singlet due to the spontaneous emission and stimulated emission by CMB photons, respectively.  On the other hand, the first and second terms on the right-hand side express the transition from singlet to triplet due to the collision and UV scattering, respectively.  The third term on the right-hand side expresses the stimulated transition from singlet to triplet by CMB photons.

According to Einstein relation, Einstein coefficients hold following relation;
\begin{equation}
A_{10} = \frac{2h\nu_{21}^{3}}{c^{2}}B_{10} 
\label{eq:A}
\end{equation}
\begin{equation}
B_{01} =3B_{10}.
\label{eq:B}
\end{equation}

As referred above Rayleigh-Jeans approximation can be applied to the CMB intensity. Thus, the CMB intensity is expressed with CMB temperature $T_{\gamma}=2.73(1+z)$ as follow,

\begin{equation}
I_{{\rm CMB}} = \frac{2\nu_{21}^{2}}{c^{2}}k_{{\rm B}}T_{\gamma}.
\label{eq:CMB}
\end{equation}

In the equilibrium state, the ratio between excitation and de-excitation rates holds the following relation with kinetic temperature $T_{\rm gas}$\citep{1958PIRE...46..240F}:

\begin{equation}
\frac{C_{01}}{C_{10}} =\frac{g_{1}}{g_{0}}\exp\left(1-\frac{T_{*}}{T_{\rm gas}}\right)= 3\exp\left(1-\frac{T_{*}}{T_{\rm gas}}\right),
\label{eq:ex_de}
\end{equation}
where $T_{{\rm gas}} \gg T_{*}=h\nu_{21}/k_{\rm B}=0.082 {\rm mK}$. 
Here, we define the color temperature of the Lyman-$\alpha$ photons $T_{C}$ via

\begin{equation}
\frac{P_{01}}{P_{10}}\equiv 3\left(1-\frac{T_{*}}{T_{\alpha}}\right)
\label{eq:color}
\end{equation}
Note that we often use a condition that $T_{\alpha}$ is coupled to kinetic temperature $T_{{\bf gas}}$ ($T_{\alpha}$=$T_{{\bf gas}}$) due to the recoiling of Lyman-$\alpha$ photons. This condition holds when there exists large number of Lyman-$\alpha$ photons \citep{1952AJ.....57R..31W,1959ApJ...129..536F,1959ApJ...129..551F}.
Now, we can rewrite eq.(\ref{eq:equilibrium}) with eq.(\ref{eq:A})-eq.(\ref{eq:color}).
\begin{eqnarray}
    \label{eq:Ts}
    T_\mathrm{S}^{-1}=
    \frac{T_{\gamma}^{-1} + x_\mathrm{c} T_\mathrm{gas}^{-1} + x_\mathrm{\alpha} T_\mathrm{\alpha}^{-1}
    }{1+ x_\mathrm{c}+ x_\mathrm{\alpha}},
\end{eqnarray}
Here, we defined coupling coefficients for collisions and scattering of Lyman-$\alpha$ scattering, $x_{c}, x_{\alpha}$ respectively;
\begin{equation}
x_{c} \equiv \frac{C_{10}T_{*}}{A_{10}T_{\gamma}} 
\label{eq:eq14}
\end{equation}
\begin{equation}
x_{\alpha} \equiv \frac{P_{10}T_{*}}{A_{10}T_{\gamma}}
\label{eq:eq15}
\end{equation}

\subsubsection{Collisional coupling}
We consider the collisional excitation and de-excitation by scattering between a hydrogen atom and other particles.  The scattering happens in dense gas. The main collision processes are (1) neutral hydrogen atom collision (H-H) (2) neutral hydrogen atom - electron collision (H-e) (3) neutral hydrogen atom - proton collision (H-p).

The coupling coefficient for species $i$ (H-H, H-e, H-p) is expressed by
\begin{equation}
x_{c}^{i} \equiv \frac{C_{10}^{i}T_{*}}{A_{10}T_{\gamma}}=\frac{n_{i}\kappa_{10}^{i}}{A_{10}}\frac{T_{*}}{T_{\gamma}},
\label{eq:coupling}
\end{equation}
where $\kappa^{i}_{10}[{\rm cm^{3} s^{-1}}]$ is the rate coefficient which describes how often collisions occur. The rate coefficient is a function of temperature. $n_{i}$ is the number density of each species. The total collisional coefficient is sum of each collisional coefficient given by 
\begin{equation}
x_{c} = x_{c}^{{\rm HH}}+x_{c}^{{\rm eH}}+x_{c}^{{\rm pH}}, 
\label{eq:collision2}
\end{equation} 

\begin{figure}[htbp]
\includegraphics[width=1.0\hsize]{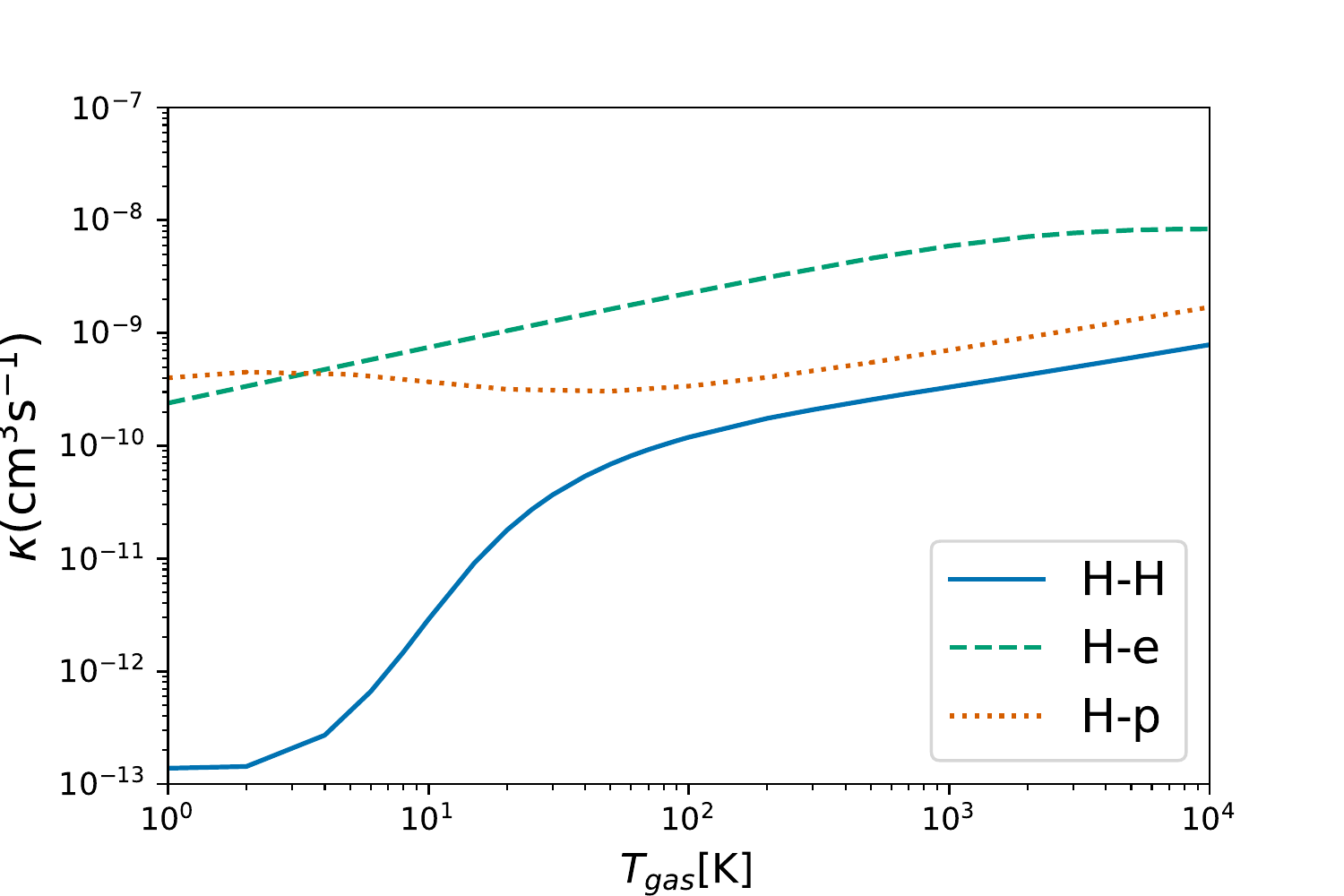}
\caption{The rate coefficients as function of temperature for each collisions, H-H collisions(solid line), H-e collisions(dashed line), H-p collisions(dotted line). Replot from \citet{2005ApJ...622.1356Z,2007MNRAS.374..547F,2007MNRAS.379..130F}.}
\label{fig:kappa}
\end{figure}

The rate coefficient can be calculated by quantum physics \citep{2005ApJ...622.1356Z, 2007MNRAS.374..547F, 2007MNRAS.379..130F}.  We show $\kappa^{i}$ for each collision in Fig.\ref{fig:kappa}.  From Fig.\ref{fig:kappa}, we can see that the rate coefficients for H-e and H-p collisions change gradually as temperature increases.  On the other hand, the rate coefficient for H-H collisions changes drastically at around $T\sim 10{\rm K}$.   This is because hydrogen atom is unable to move violently due to their heavy mass and small cross-section at $ T\lesssim 10{\rm K}$.  Before the EoR, the neutral hydrogen atom is the dominant component, thus H-H collision is the main process. Once the EoR started, the number of electrons gradually increases. Thus, the collision between hydrogen atom and electron starts to play an important role.  As we describe later, the kinetic temperature of the IGM drastically increases and becomes larger than $10^{4}{\rm K}$ after X-ray heating turns on. At that time the collision between the hydrogen atom and electron becomes most dominant.

\subsubsection{Wouthyusen-Field effect}

\begin{figure}
    \includegraphics[width=1.0\hsize]{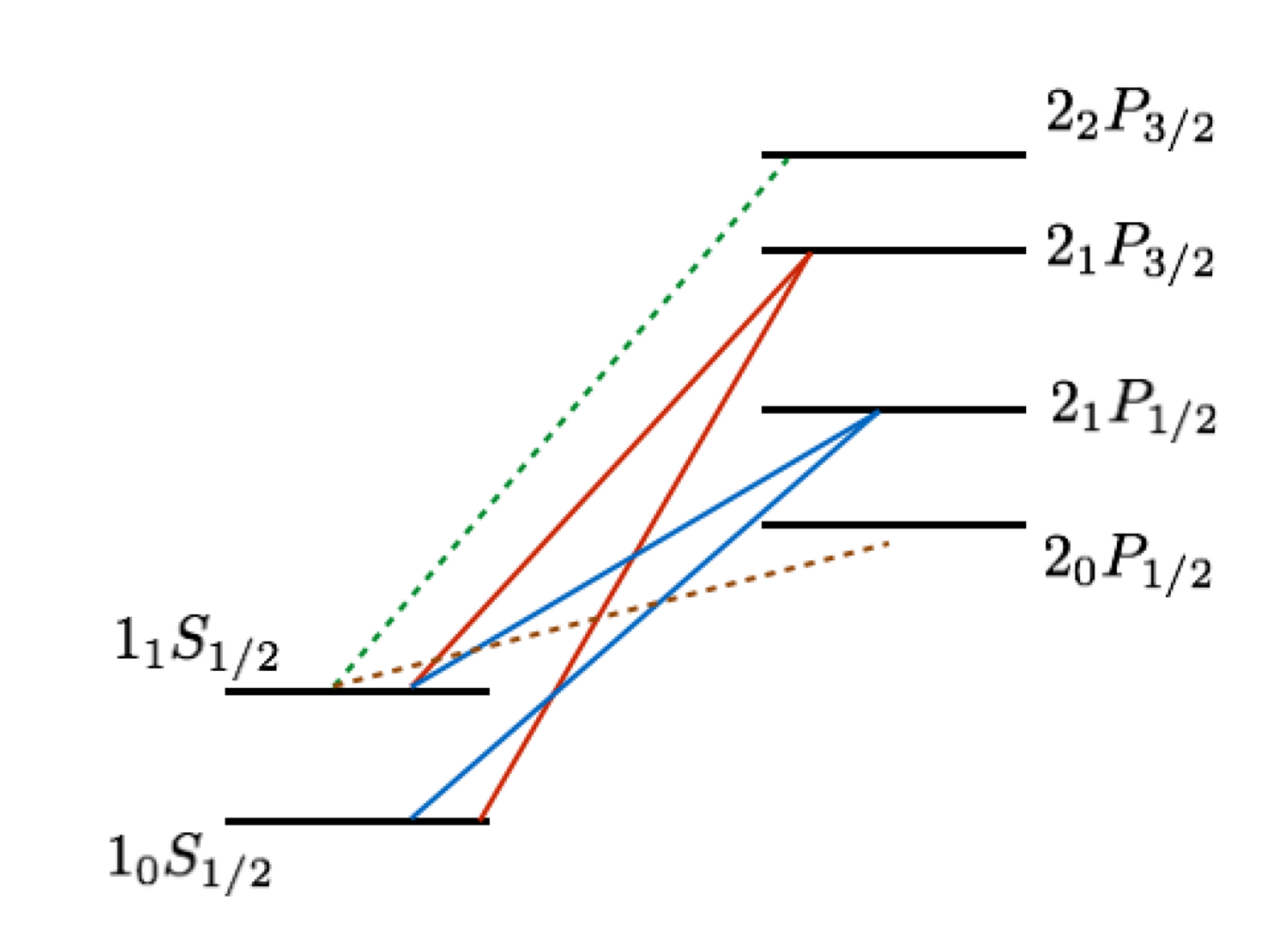}
    \caption{Hyperfine structure of the hydrogen atom. The transition between S states and P states is related to the Wouthuysen-Field effect. Solid lines show the process allowed by the selection rule and spin-flip can occur. While dashed lines are allowed, these processes do not contribute to spin-flip.}
    \label{fig:WF}
\end{figure}

We introduce an important physical mechanism related to spin temperature. Resonant scattering of Lyman-$\alpha$ photons emitted by first stars provides paths of energy transition between singlet and triplet in the neutral hydrogen atom. This process is known as ``Wouthyusen-Field (WF) effect'' \citep{1952AJ.....57R..31W,1959ApJ...129..536F,1959ApJ...129..551F}. We show the schematic picture in Fig.\ref{fig:WF}. A hydrogen atom in the singlet state $1_0 S_{1/2}$ is excited to 2P states by absorbing Lyman-$\alpha$ photons. Some energy transitions from 2P states to triplet state $1_1 S_{1/2}$ are allowed by re-emission of Lyman-$\alpha$ photons (Some paths are not allowed by the selection rule in quantum mechanics). Hydrogen atoms can change their energy state between singlet and triplet through absorption and re-emission of Lyman-$\alpha$ photons. As described in the WF effect, Lyman-$\alpha$ photons play an important role to exchange energy between singlet and triplet in a neutral hydrogen atom. \citet{1959ApJ...129..536F} showed that the WF effect leads to a coupling of the color temperature to kinetic temperature ($T_{\alpha} \sim T_\mathrm{gas}$) under the condition where there is a large number of Lyman-$\alpha$ photons.

We here revisit coupling coefficient by Lyman-$\alpha$ scattering shown in eq.\ref{eq:eq15}. The eq.\ref{eq:eq15} is defined by de-excitaion rate $P_{10}$. This coupling coefficient is re-written with scattering rate of Lyman-$\alpha$ photon, $P_{\alpha}$, as follow\footnote{ You can find detail derivation of $\frac{4}{27}$ in \url{https://casper.astro.berkeley.edu/astrobaki/index.php/Wouthuysen_Field_effect}} \citep{1958PIRE...46..240F,1959ApJ...129..536F}:
\begin{equation}
x_{\alpha} = \frac{4P_{\alpha}T_{*}}{27A_{10}T_{\gamma}}.
\end{equation}

The scattering rate $P_{\alpha}$ is expressed by 
\begin{equation}
P_{\alpha}=4\pi \chi_{\alpha}\int d\nu J_{\nu}(\nu)\phi_{\alpha}(\nu)
\label{eq:Palpha}
\end{equation}
where $\chi_{\alpha}\equiv(\pi e^{2}/m_{e}c)f_{\alpha}, f_{\alpha}=0.4162$ is the oscillation strength of Lyman-$\alpha$ transition and $\phi_{\alpha}(\nu)$ is the line profile for Lyman-$\alpha$ absorption.  $J_{\nu}(\nu)$ is angle averaged specific intensity of the background radiation field. For example, \citet{2004ApJ...602....1C} and \citet{2006MNRAS.367..259H} discussed the detail treatment of $P_{\alpha}$.

\subsubsection{Excess radio background}

In the last part of this section, as the scenario which affects the spin temperature, we introduce an excess radio background scenario. The excess radio background is one of the scenarios to explain absorption feature detected by EDGES\citep{2019MNRAS.486.1763F,2020MNRAS.499.5993R}. Astrophysical sources such as accreting supermassive blackholes or supernovae are candidates of excess radio background sources. They generate excess radio background via synchrotron emission produced by electron accelerated by magnetic fields. Interestingly, before EDGES reported the deep trough at 78MHz, ARCADE2 instrument has shown the evidence of excess radio background above CMB at low frequency\citep{2011ApJ...734....5F} and recently the excess radio background is confirmed by LWA1 experiment at 40-80 MHz\citep{2018ApJ...858L...9D}. Thus, excess radio background scenarios is not exotic scenario, but one of the possible scenarios to explain EDGES's result. In the excess radio background scenario, total radio background has the form of sum of CMB temperature and excess radio as shown by

\begin{equation}
T_{\mathrm{rad}}=T_{\mathrm{CMB}}(1+z)\left[1+A_{\mathrm{r}}\left(\frac{\nu_{\mathrm{obs}}}{78 \mathrm{MHz}}\right)^{\beta}\right]
\end{equation}
where $\nu_{obs}$ is the observed frequency, $A_{r}$ is the amplitude defined relative to CMB temperature and $\beta$ is the spectral index. By taking excess radio background into account, the spin temperature formalism shown by eq.\ref{eq:Ts} is modified. Thus, excess radio background affects not only the local spin temperature, but also the 21cm brightness temperature.

\subsection{Thermal history}
So far, we introduced the spin temperature and described coupling coefficients. To calculate the spin temperature, we also need to know how the gas kinetic temperature $T_{{\rm gas}}$ and the color temperature $T_{\alpha}$ evolve. However, as we mentioned before, the color temperature is coupled to gas kinetic temperature under the condition  in the large number of Lyman-$\alpha$ photons. Thus, we just need to know how the gas kinetic temperature evolves. The evolution of gas kinetic temperature is described by\footnote{This equation comes from first law of thermodynamics} \citep[see e.g.][]{2012RPPh...75h6901P}

\begin{equation}
\frac{\mathrm{d} T_{\mathrm{gas}}}{\mathrm{~d} t}=\frac{2 T_{{\mathrm{gas}}}}{3 n} \frac{\mathrm{d} n}{\mathrm{~d} t}+\frac{2}{3 k_{B}} \sum_{j} \frac{\epsilon_{j}}{n}.
\label{eq:gas}
\end{equation}

Here,$n$ is the number density of gas particles, and $\epsilon_{j}$ is the heating rate per unit volume for the process $j$. The first term accounts for adiabatic cooling of the gas due to cosmic expansion and the second term accounts for the heating or cooling process. As heating mechanisms, we mainly have 3 processes. (1) Compton heating, (2) X-ray heating, (3) Lyman-$\alpha$ heating. At high redshift before first nonlinear objects emerge, the Compton heating caused by scattering between CMB and residual free electron is the dominant heating mechanism. As redshift is decreasing, the Compton heating becomes ineffective. Once nonlinear objects are formed, X-ray photons emitted from these objects heat IGM. X-ray heating is the most important source of energy injection into the IGM. As the source of X-ray photons. The IGM gas is first photoionized by X-ray photons and this generates energetic photo-electrons. These photo-electrons distribute their energy into the IGM by collision with the HI atom in the IGM.  The X-ray photon with energy $E$ has a long comoving mean free path as shown\citep[e.g.][]{2006PhR...433..181F}

\begin{equation}
    \lambda_{\mathrm{X}}\sim 4.9 \bar{x_{\mathrm {HI}}}^{-1/3}\left(\frac{1+z}{15}\right)^{-2}\left(\frac{E}{300\mathrm{eV}}\right)^3 \mathrm{Mpc}.
\end{equation}

Therefore, X-ray photons can heat the gas far from X-ray sources. In nearby star-forming galaxies, high mass X-ray binaries (HMXBs), which are X-ray binaries fed by the winds of massive companion, are the main contributor to X-ray luminosity\citep[e.g.][]{2004MNRAS.347L..57G,2012MNRAS.426.1870M}. Thus, HMXBs are also expected to be reliable X-ray sources at high redshift\citep[e.g.][]{2014Natur.506..197F,2014MNRAS.440.3778J,2014ApJ...791..110X}. We often extrapolate correlation between the star formation rate(SFR) and the X-ray luminosity $L_{\mathrm{X}}$ correlation found in local HMXBs to high redshift with unknown renormalization factor $f_{\mathrm{X}}$.
\begin{equation}
    L_{\mathrm{X}}=3\times 10^{40}f_{\mathrm{X}}\left(\frac{\mathrm{SFR}}{M_{\odot}\mathrm{yr}^{-1}}\right)\mathrm{erg s^{-1}}.
\end{equation}
\citet{2021arXiv210807282T} implies that the $L_{\mathrm{X}}/\mathrm{SFR}$ at high redshift is consistent with expectation of local HMXB in metal-poor environment. In addition to HMXBs, stellar objects in galaxies and quasars (and also supermassive blackholes) are also candidates of the source of heating\citep[e.g.][]{2017MNRAS.468.3785R,2019MNRAS.487.1101R,2018MNRAS.476.1174E,2022PhRvD.106d3539K}. Roughly, the second term in eq.\ref{eq:gas} by X-ray hearing is $ (2\epsilon_{{\rm X}}/3k_{{\rm B}} n H(z)) \sim 10^{3} f_{{\rm X}} \mathrm{~K}$ \citep[e.g][]{2006PhR...433..181F}.

In addition that Lyman-$\alpha$ photon couples the spin temperature to the kinetic temperature by scattering, resonant scattering of Lyman-$\alpha$ photons also heat the gas through atomic recoil scattered by Lyman-$\alpha$ photons. The second term of the eq.\ref{eq:gas} contributed from the Lyman-$\alpha$ heating is roughly $ 2\epsilon_{\alpha}/(3H n_{{\rm H}} k_{{\rm B}} T_{{\rm gas}}) \approx 0.80T_{{\rm gas}}^{-4 / 3}(10/1+z)$ when WF effect is effective \citep[e.g][]{2006PhR...433..181F}. Thus, the heating by the Lyman-alphas is negligibly small compared with the X-ray heating. This is because the Lyman-$\alpha$ heating is nearly fully compensated by cooling. The gas energy obtained by recoil from Lyman-$\alpha$ photons is back to the Lyman-$\alpha$ photons because some of the photons are scattered to the blue side of the Lyman-$\alpha$ background profile and gain energy from the gas \citep[See more detail in ][]{2004ApJ...602....1C,2006PhR...433..181F}. On the other hand, some works noted that the number of photons used for heating and back to photons from the gas is not similar\citep{2006MNRAS.372.1093F,2006MNRAS.367..259H,2007ApJ...655..843C}. In that case, Lyman-$\alpha$ photons can heat the gas to a temperature of $\sim$ 100 K\citep{2007ApJ...655..843C}. Recently, \citet{2021MNRAS.506.5479R} evaluated the effect of Lyman-$\alpha$ heating taking multiple scattering of Lyman-$\alpha$ photons. They also showed that gas temperature reaches $\sim$ 100 K at $z=6$ if we account for Lyman-$\alpha$ heating, although it is around a few K if we do not account for Lyman-$\alpha$ heating. They conclude that Lyman-$\alpha$ heating and CMB heating become important when X-ray heating is inefficient ($f_{\mathrm{X}} \lesssim 0.1$ for a SED of X-ray binaries).

As another heating process, the CMB may play an important roles. The energy of CMB is transferred to Lyman-$\alpha$ photons. The Lyman-$\alpha$ photons heat the gas via atomic recoils. Thus, in the absence of X-ray heating, the mechanism of energy transfer between CMB photons and Lyman-$\alpha$ photons is worth considering \citep[e.g.][]{2018PhRvD..98j3513V}. However, we also note that there is a  controversy about CMB heating \citep[e.g.][]{2021RNAAS...5..126M}.

\begin{figure}
    \includegraphics[width=1.0\hsize]{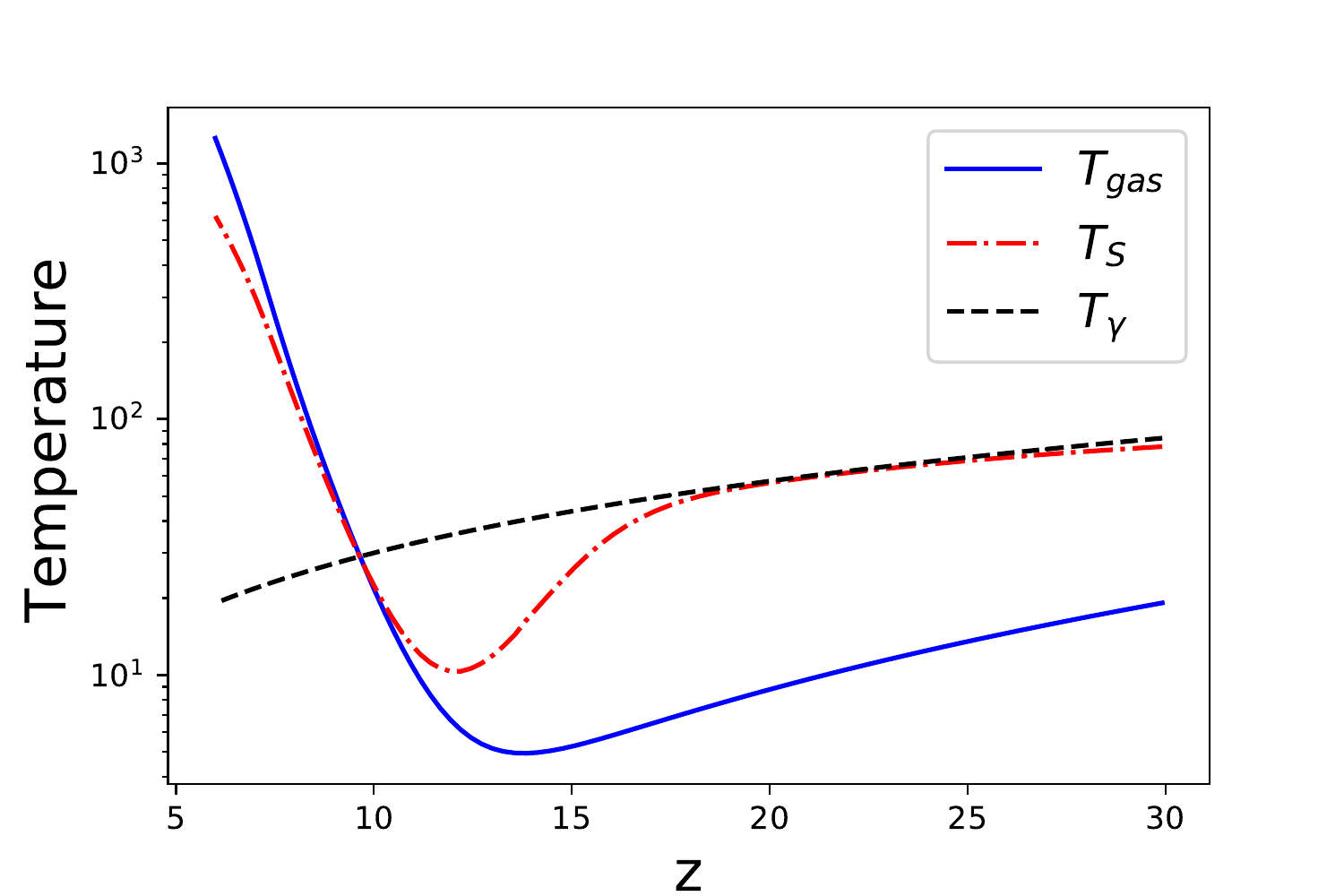}
    \caption{Time evolution of temperatures. We show the evolution of kinetic temperature (blue solid line), spin temperature (red dash-dot line) and CMB temperature(black dashed line).}
    \label{fig:temperature}
\end{figure}

In Fig.\ref{fig:temperature}, we show thermal evolution of each temperature by using 21cmFAST with default parameter set \citep{2011MNRAS.411..955M,2019MNRAS.484..933P}.  Note that this is the particular result of the model used in 21cmFAST. We refer to the 21cmFAST in section \ref{sec:model}. We here explain the behavior of temperatures. Due to cosmic expansion, the kinetic temperature ($T_{\mathrm{gas}}$) decreases adiabatically as redshift decreases. Once heating by high energetic photons emitted by compact objects such as X-ray binaries becomes effective, the kinetic temperature starts to increase drastically. The spin temperature ($T_{\mathrm{S}}$) is coupled with CMB temperature ($T_{\gamma}$) at high redshift before first luminous objects form by the collision between CMB photons and neutral hydrogen atoms. Once the first luminous objects formed and emitted Lyman-$\alpha$ photons, the WF effect becomes effective. Thus, the spin temperature is coupled to the kinetic temperature and thermally evolves together.


\subsection{Global signal}
\begin{figure}[htbp]
\includegraphics[width=1.0\hsize]{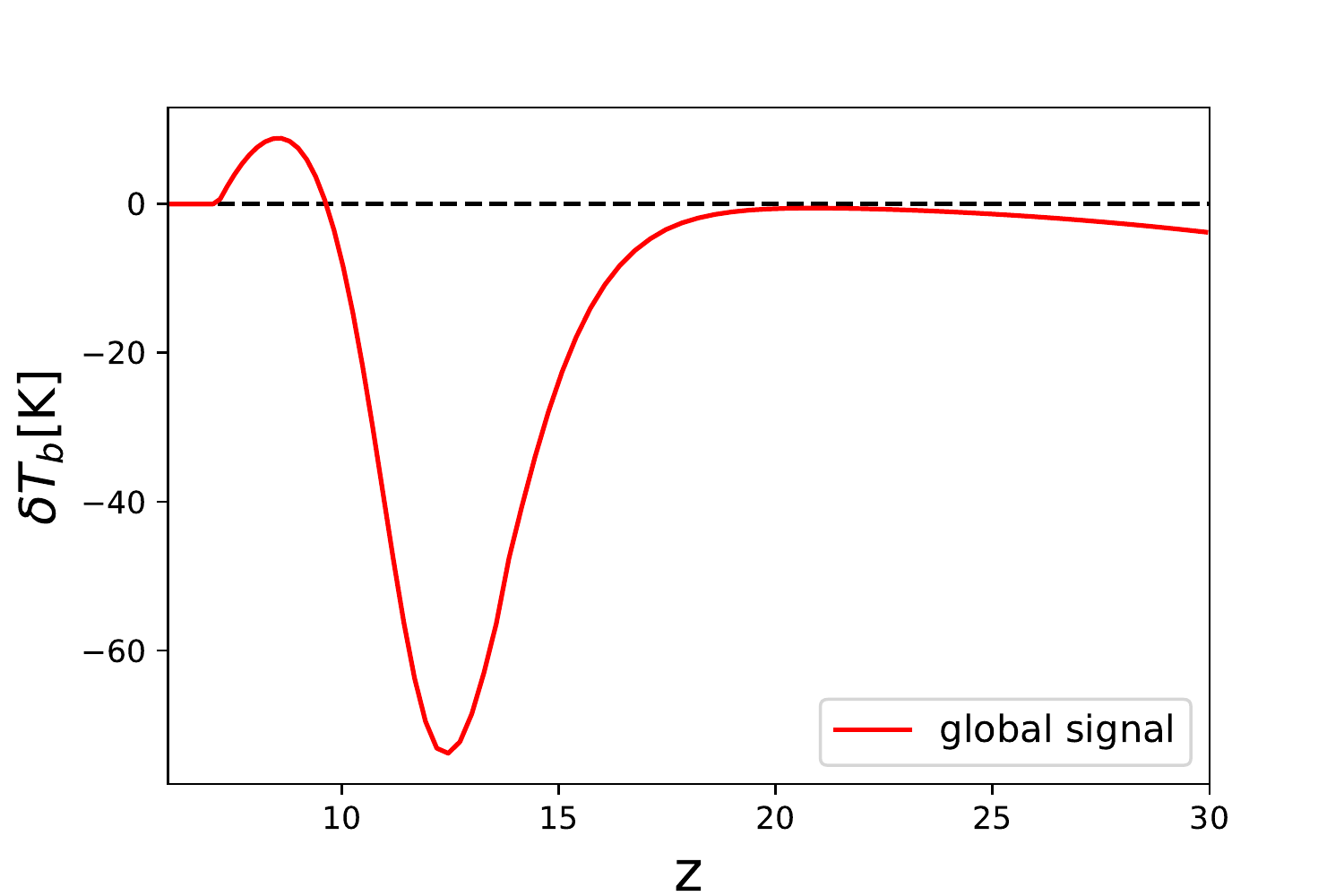}
\caption{The time evolution of global signal.}
\label{fig:global}
\end{figure}

The sky-averaged brightness temperature, called 21cm global signal, is also a key quantity of the 21cm line signal. As similar as Fig.\ref{fig:temperature}, we also plot 21cm global signal by using 21cmFAST in Fig.\ref{fig:global}. The 21cm global signal mainly traces the behavior of the spin temperature and neutral hydrogen atom fraction. When the spin temperature is close to the CMB temperature(at $z \gtrsim 17$), the global signal is close to zero. Once the WF effect turns on, the spin temperature becomes below the CMB temperature and approaches kinetic temperature, thus the global signal becomes negative. After X-ray heating becomes effective and the spin temperature is above the CMB temperature, the global signal becomes positive and it produces a deep trough at $z\sim 13$. When the spin temperature becomes enough higher than the CMB temperature, the global signal does not depend on the spin temperature because it is saturated (see eq. \ref{eq:brightness}) and ionization history determines the global signal. Since the neutral fraction is decreasing at the EoR, the global signal is also decreasing at the EoR and finally becomes zero.

\subsection{Modeling the 21cm signal}\label{sec:model}

Cosmic 21cm signal is deeply related to the underlying astrophysics driving reionization and cosmic dawn. To interpret the cosmic 21cm signal properly, we need to model such astrophysics. For this purpose, we have two types of simulations, direct full-numerical simulations, and semi-numerical simulations.

Full-numerical simulations are designed to investigate underlying astrophysics most accurately by solving basic equations. The full-numerical simulations consist of the dynamics of baryons and dark matter and the radiative transfer of photons responsible for ionizing or heating of the IGM. 

Radiation hydrodynamics (RHD) simulations, in which the radiative transfer is coupled with hydrodynamics, consistently solve the evolution of the ionization field and the structure formation
\citep[e.g.][]{1997ApJ...486..581G,2000MNRAS.314..611C,2003MNRAS.344L...7C,HS2013,2014MNRAS.442.2560W,2014ApJ...789..149S,2016MNRAS.463.1462O,2020MNRAS.496.4087O,2018MNRAS.479..994R,2022MNRAS.511.4005K}. 
The difficulty in RHD simulations is their expensive computational cost. 
To simulate the representative evolution of the Universe, a sufficiently large simulation box size (several $100\mathrm{Mpc}$) is required. 
Furthermore, high mass/spatial resolution is important to resolve all ionizing sources including mini-halos.  
The computational cost for post-processing radiative transfer simulations is relatively cheap but is still more expensive than semi-numerical simulations described later\citep[e.g.][]{2006MNRAS.369.1625I,2007ApJ...671....1T,2014MNRAS.439..725I,2018PASJ...70...55I, 2020MNRAS.491.1600M}. 
In general, it is impossible to satisfy both requirements, 
thus modeling of ionizing sources is crucial even in the full-numerical simulations.

Rather than simulating all astrophysical processes performed by full-numerical simulations, we have another choice called semi-numerical simulation. In semi-numerical simulations, we make a number of simplifying approximations in astrophysical processes to reduce computational costs. With simplifying approximation in astrophysical process at small scales, semi-numerical simulations can achieve large dynamical range, that is to say, we can simulate the universe from small scales to large scales. The 21cmFAST is one of the semi-numerical simulation\citep{2011MNRAS.411..955M,2019MNRAS.484..933P}. In the 21cmFAST, their treatment of ionization is based on excursion-set formalism and bypass radiative transfer by replacing it with an approximation in which we count the number of ionizing photons and compare it with recombination\citep[e.g.][]{2007ApJ...669..663M,2010MNRAS.406.2421S,2011MNRAS.411..955M,2014MNRAS.440.1662S,2019MNRAS.484..933P}. In \citet{2019MNRAS.484..933P}, they introduced flexible and physical parameters motivated by high redshift galaxy properties and implemented it in the latest version of 21cmFAST. They model the star formation rate and ionizing escape fraction by scaling with masses of their host dark matter halos, and directly compute inhomogeneous recombination with sub-grid model. They calibrated their model by using current observations of rest-frame UV luminosity function at high redshift.

\subsection{21cm line power spectrum}\label{sec:21cmps}
To extract astrophysical and cosmological information from cosmic 21cm line signal, we need to interpret observational 21cm line signal. As one of the simplest methods to exploit fruitful information from cosmic 21cm line signal, we often use 21cm line power spectrum. The 21~cm power spectrum characterizes the fluctuations in the 21~cm brightness temperature. The 21~cm power spectrum is defined by \citep[e.g.][]{fur} 
\begin{equation}
  \langle \delta T_b(\mathbf{k}) \delta T_b(\mathbf{k}^{\prime})\rangle
= (2\pi)^3 \delta(\mathbf{k}+{\mathbf{k}^{\prime}}) P_{21}(k).
\end{equation}
 
We often use the normalized 21~cm power spectrum, $k^{3}P_{21}(k)/2\pi^{2}$. Note that the normalized 21cm power spectrum does not have a length dimension, but has the dimension of temperature. 

\begin{figure}
    \includegraphics[width=1.0\hsize]{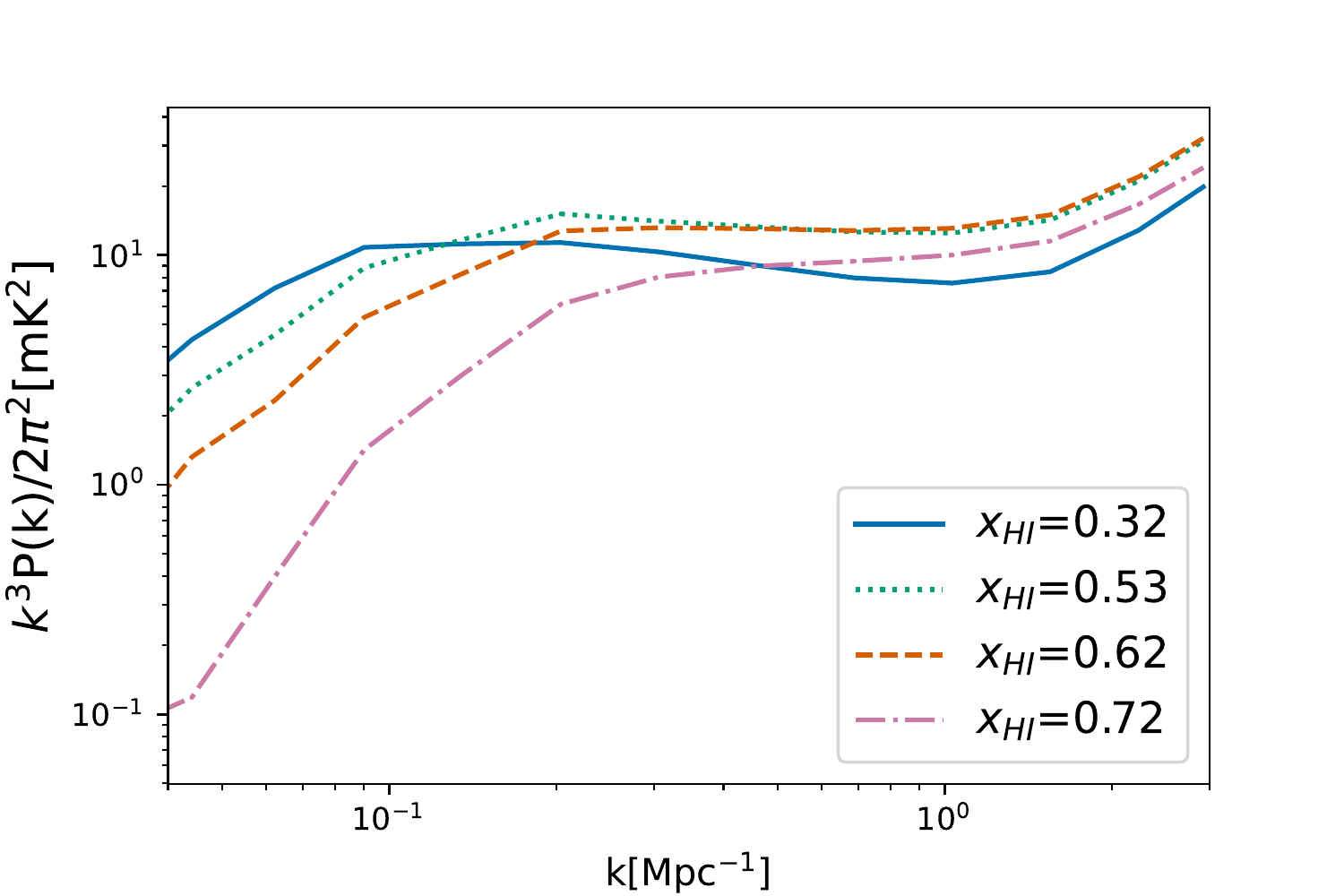}
    \includegraphics[width=1.0\hsize]{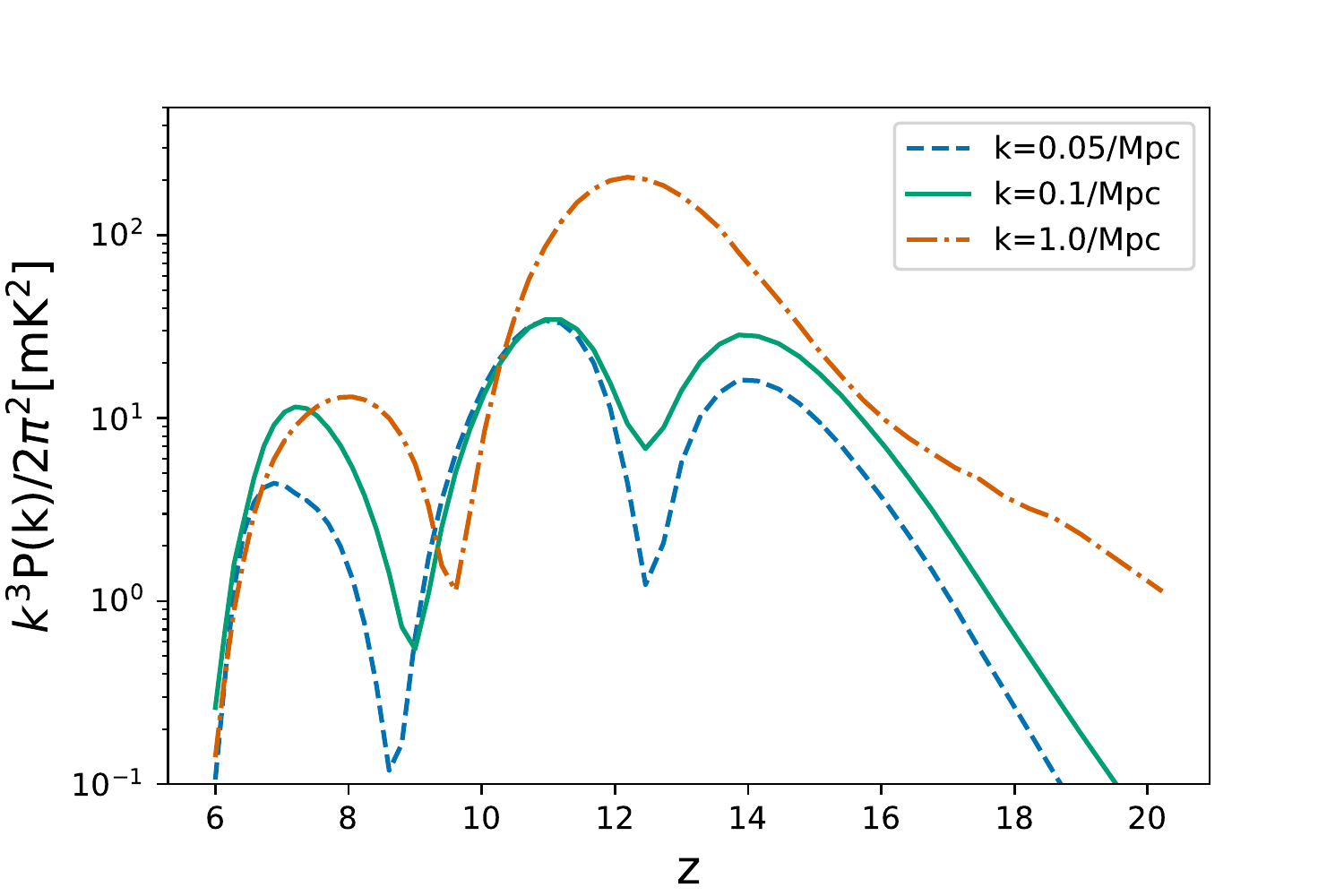}
    \caption{({\it Top}) The 21cm line power spectrum as function of wavenumber at different neutral hydrogen fraction. ({\it Bottom}) The 21cm line power spectrum as function of redshift at fixed wavenumbers.}
    \label{fig:21cmps}
\end{figure}

In Fig.\ref{fig:21cmps}, we show the 21cm line power spectrum. As similar as Fig.\ref{fig:temperature}, we calculate the power spectrum by 21cmFAST with default parameter set.  At the top panel of Fig.\ref{fig:21cmps}, we show the 21cm line power spectrum as a function of wavenumbers at different neutral hydrogen fractions. We can see the bumps in each 21cm line power spectrum around 0.1-0.3$\mathrm{Mpc}^{-1}$. These bumps correspond to a typical size of ionized bubbles. We can also see that the amplitude of the 21cm line power spectrum at $k \gtrsim 0.5\mathrm{Mpc^{-1}}$ becomes smaller as a neutral fraction of hydrogen atom becomes smaller. This indicates that the 21cm line power spectrum at smaller scales is dominated by the fluctuation of the neutral hydrogen atom. At the bottom of Fig.\ref{fig:21cmps}, we show the redshift evolution of the 21cm line power spectrum at fixed wavenumbers. We can see characteristic peaks in the 21cm line power spectrum as a function of redshift. For $k=0.05, 0.1\mathrm{Mpc}^{-1}$, we can see three peaks. On the other hand, we see only 2 peaks for $k=1.0 \mathrm{Mpc}^{-1}$. In section \ref{sec:moment}, we revisit the reason why the number of peaks differs depending on the wavenumber scales in more detail. In this subsection, we only explain each peak that appeared in the 21cm line power spectrum in the case of $k=0.05, 0.1\mathrm{Mpc}$. 

Each peak corresponds to astrophysical effects. This means that we can know the period when astrophysical effects become effective utilizing the 21cm line power spectrum as a function of redshift. The peak at $z\sim 14$ is due to the Wouthuysen Field (WF) effect. From Fig.\ref{fig:temperature}, we can understand this is due to the WF effect. The WF effect becomes effective at $z\sim 15$. The peak at $z \sim 11$ is due to X-ray heating. We also can see that X-ray heating becomes effective at $z \sim 11$ in Fig.\ref{fig:temperature}. The trough corresponding to X-ray heating generates a characteristic peak in the 21cm line power spectrum. The peak that appeared at $z \sim 7$ is generated by the reionization. Note that the behaviors of the 21cm power spectrum shown in Fig\ref{fig:21cmps} are for the specific case. However, for most standard astrophysics and cosmology models, we can find similar behavior in the 21cm line power spectrum.

\section{21cm Signal from Pop III stars}

In the previous section, we introduced basic physics of the 21cm line. In this section, we look more in detail at the physics of the 21cm line. One of the astrophysics which affects the 21cm line signal is Pop III star, which is the first generation of stars formed in early universe. We focus on how Pop III stars impact on the 21cm line signal. Pop III stars are theoretically predicted to form in mini-halos (MHs) at high redshifts. 
Since the Pop III stars in MHs are thought to be faint compared to typical high-$z$ galaxies, direct detection of their stellar light is almost impossible even with the next generation telescopes such as JWST. 
However indirect detection through the 21cm brightness temperature, which is affected by the stellar radiation, could be possible.  
Therefore it is expected that we obtain the information of the Pop III stars from observed 21cm signals. 
For the purpose we need to understand the relation between the properties of Pop III stars and the expected 21cm signal. 

\subsection{21cm signal around individual MHs}
\label{subSec:individual}

We first focus on 21cm signal around individual MHs. \cite{CHen2008} have calculated the 21-cm signature around a MH. Since Pop III stars have high effective temperature compared to galaxies composed of Pop II stars, the 21cm signal around Pop III stars shows characteristic signature. 
In the most vicinity of a Pop III star, 21cm signal does not appear because of high ionized fraction. The size of this region is basically determined by the number of ionizing photons from the Pop III star. 
Just outside the non-signal region, the gas is partially ionized and heated above the CMB temperature. As a result, 21cm signal appears as emission. 
The region outside the emission region show 21cm absorption feature, since the spin temperature is tightly coupled with the cold IGM via the strong WF coupling. 
In the most distant region from the Pop III stars, there is no remarkable feature because the spin temperature almost corresponds to the CMB temperature.  

As mentioned above, the size of the 21cm signal region is determined by the WF effect. Therefore it is important to solve the radiative transfer (RT) of Lyman-$\alpha$ photons. 
\cite{Yajima2014} have also calculated the 21cm signal around Pop III stars, solving the RT of Lyman-$\alpha$ photons to evaluate the WF coupling correctly. The computed 21cm signal distributions indicated that the size of the individual signal is too small to detect even with the SKA if a MH hosts a Pop III star with $\sim 10^2\rm M_{\odot}$. 

\cite{TT2018} have further improved this study by focusing on the stellar and halo mass dependence of the 21cm signal. 
They have performed RHD simulations, in which MHs can be spatially resolved. 
Such RHD simulations allow us to consider the appropriate escape fraction and the dynamical expansion of HII region. 
They showed that the size of 21cm signal region strongly depends on the stellar and halo mass since the escape fraction depends on them. They also found that the dynamical expansion of HII region hardly affects the size of 21cm signal region.

In short, these studies have revealed that the 21cm signal indeed reflects the properties of Pop III stars, but the individual spatial distribution of the signal cannot be detected even with the SKA. 
Therefore, in order to assess the properties of Pop III stars, it is important to focus on the mean value like the global signal and/or statistical values of the 21cm line.

\subsection{Redshift evolution of the global 21cm signal}\label{subSec:globalsignal}

The global signal of the 21cm line can reflect the statistical properties of star formation in the early Universe through UV and X-ray radiation. While recent observations of galaxies have revealed the star formation history up to redshift $\sim 7$, the star formation history above then is not well constrained due to the sensitivity limitations of current observations. Since the cosmic reionization is now suggested to begin at redshifts above 7 \citep{Planck2020}, it is important to understand the star formation history at the epoch. In particular, little is known about Population III stars due to the lack of direct observations.

Current theoretical models predict that Population III stars are typically massive unlike local metal-rich stars because of the high gas accretion rate onto a proto-star \citep[e.g.,][]{Bromm2002, Omukai2003, Yoshida2008}. Therefore, they can be efficient sources of metal enrichment in the early Universe through frequent supernova explosions. In addition, some of Population III stars could form binary systems \citep{Turk2009}, and their subsequent evolution into black hole binary systems might result in gravitational wave events in the local Universe. Thus, it is extremely important to obtain information not only on Population II stars in high-redshift galaxies, which are thought to be main cosmic reionization sources \citep{Yajima2011, Yajima2014b}, but also on Population III stars that form in mini-haloes. Recently, the global signal of 21cm-line during the epoch of cosmic reionization was observed by EDGES \citep{Bowman2018}, showing a deep absorption signal at redshift $\sim 17$. Because the typical halo mass in this epoch is small, UV radiation from Population III stars in mini-halos could induce the signal by changing the spin temperature of the hydrogen in the IGM. However, the UV photon density sensitively depends on the formation rate density of Population III stars and on the initial mass function, which has been poorly understood.

\begin{figure}
	\begin{center}
		\includegraphics[width=8cm]{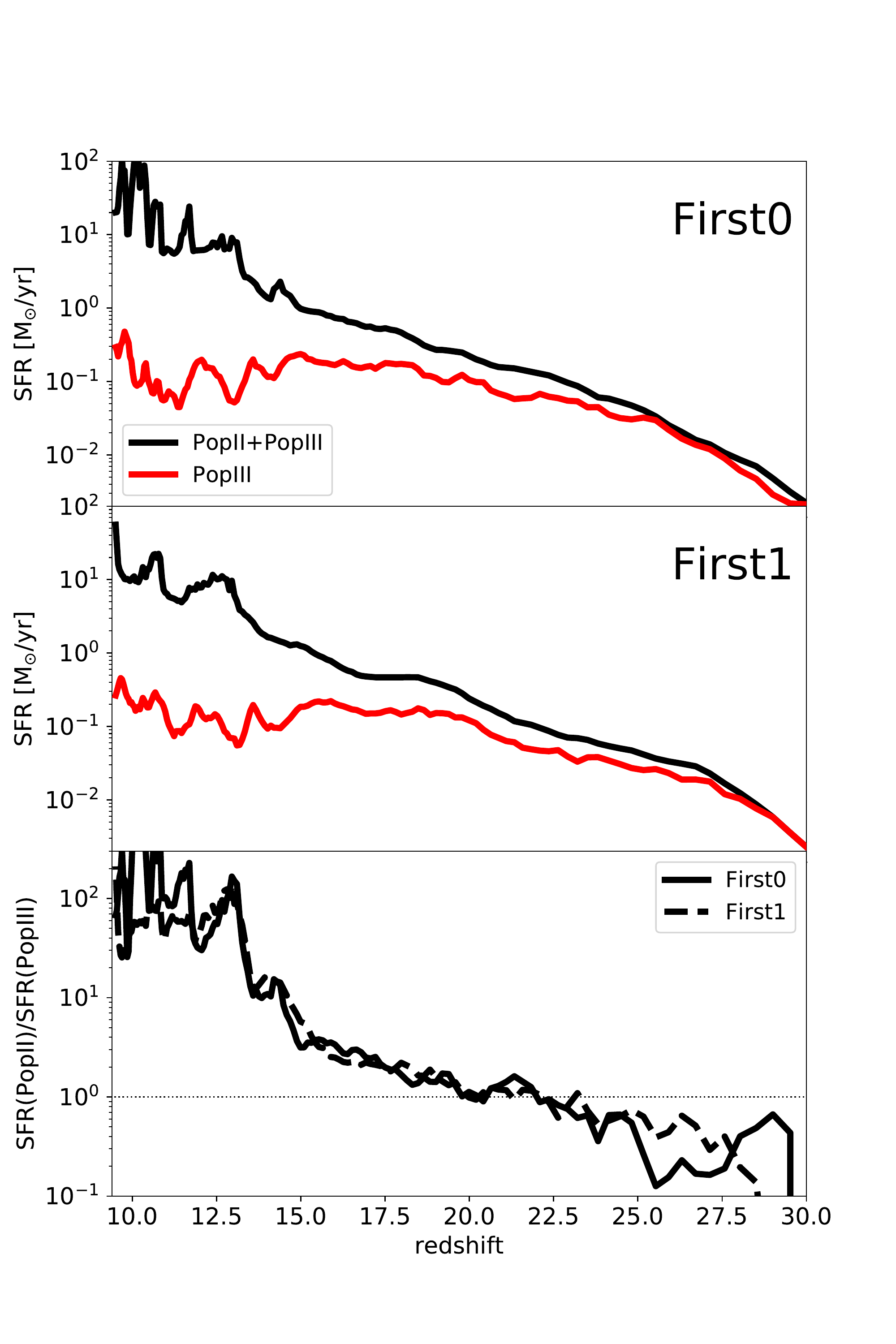}
	\end{center}
	\caption{
	
	Star formation rate histories of Population III and II stars. Red and black lines represent the star formation rates of Population III stars alone and total ones, respectively.
	Bottom panel shows the ratio of star formation rates of Population II to Population III stars. Adapted from \citet{Yajima2022}, by permission of Oxford University press on behalf of the Royal Astronomical Society.
	
		 }
	\label{fig:sfr_first}
\end{figure}

Recent large-scale cosmological hydrodynamic simulations are modeling the formation rate density of Population III stars \citep[e.g.,][]{Johnson13, Abe2021}. \citet{Yajima2022} suggested that the transition of the main stellar population depended on the place in the large-scale structure. They showed, in the case of high-density regions, the transition from Population III to Population II occurred at redshifts of $\gtrsim 20$ due to the early metal enrichment via the active star formation as shown in Figure~\ref{fig:sfr_first}. The initial mass function of Population III stars is also becoming clearer with recent numerical simulations \citep{Susa2014, Hirano2015, Sugimura20, Latif2022}. These simulations show that the mass growth of massive stars is suppressed by radiation feedback 
\citep[see also,][]{Hosokawa2016}. \citet{Hirano2015} suggested the log-normal shape of the initial mass function with the mass range from $\sim 10$ to $\sim 1000~M_{\odot}$. On the other hand, other simulations proposed the different shapes of the initial mass function, and hence it is still under debate. In particular, it is difficult to follow fragmentations of circumstellar disks and their subsequent evolution for a long time due to the limitation of computational resources. Therefore, the number of the fragments changes with the conditions of the numerical simulations \citep{Susa2019}. Thus, even with the state-of-the-art simulations, the statistical properties of Population III stars are still unclear. Therefore, it is extremely important to investigate Population III and Population II stars in the early Universe using observation of the global signal of 21cm line.

\begin{figure}
\includegraphics[scale=0.4]{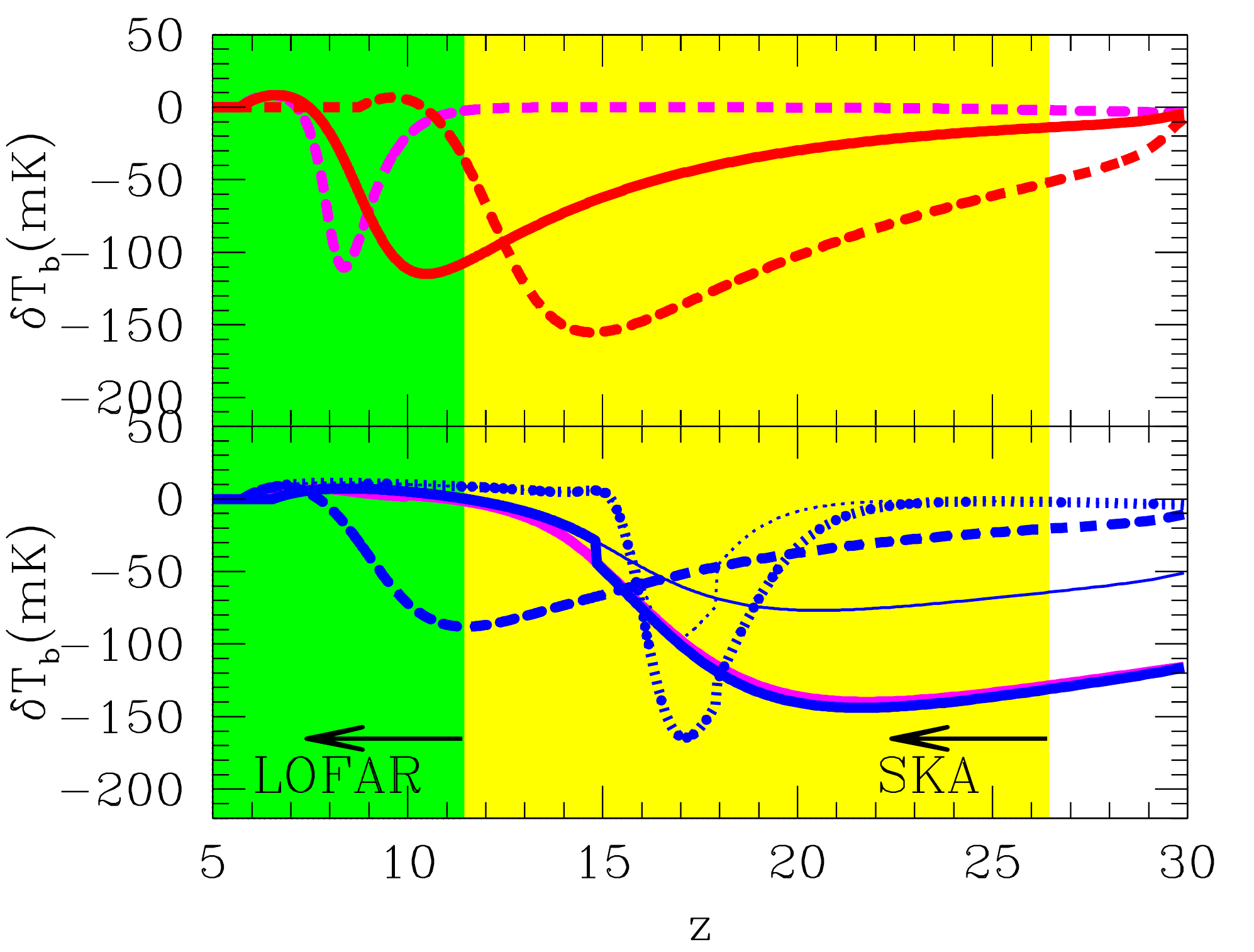}
\caption{
Redshift evolution of differential brightness temperature $\delta T_{\rm b}$ for different models . Red lines show the models without Population III stars, while Blue lines contain them.
Different line types represent the different models of star formation histories of Population III and II stars. 
Blue thin solid and dot lines represents the $\delta T_{\rm b}$ without $\rm Ly\alpha$ photons from ISM.
Shaded regions show the  range covered by LOFAR ($z \lesssim 11.4$) and SKA ($z \lesssim 26.4$).
Adapted from \citet{Yajima2015}, by permission of Oxford University press on behalf of the Royal Astronomical Society).
}
\label{fig:tb}
\end{figure}

The study of the 21cm global signal is well suited to investigate the statistical properties of Population III and II stars in the early Universe. For example, \citet{Furlanetto06a} studied the relationship between the global signal and stellar populations. According to their calculations, Population III stars decrease the absolute values of the negative signal of the differential brightness temperature by heating IGM efficiently. \citet{Pritchard2010} studied the parameter dependence of the 21cm global signal on Lya and the X-ray background and suggested future 21-cm observations will be able to constrain the parameters of Population III stars.

\citet{Yajima2015} considered several parameters for the redshift dependence of the star formation rate density of Population III stars and investigated their impact on the 21cm line global signal. Figure \ref{fig:tb} shows that the redshift dependence of the global signal changes significantly with the star formation rate densities of Population III stars. Red lines show the models without Population III stars, and blue lines do the models with Population III stars. The blue dotted line shows the Population III star formation rate density with a large peak at redshift 15. The 21cm line signal shows a deep absorption signal of $\sim -170~\rm mK$ before the peak of the star formation rate. At redshift 15, the absorption signal becomes much smaller due to the ionization and heating from Population III stars. The solid blue line considers a gradual increase and decrease in the star formation history. In this case, the absorption signal deeper than $-50~\rm mK$ can be observed over redshifts $\sim 17-30$. In addition to these studies, \citet{Yajima2015} also investigated the relationship between the initial mass function of Population III stars and the 21cm line signal. Thus, they suggested that the shape of the redshift dependence of the 21cm global signal above redshift 15 may provide constraints on the star formation history of population III stars.

Recently, \citet{Qin2020} have semi-analytically modeled the large-scale structure of the 21cm signal using a 21CMFAST code \citep{Mesinger2011}. Their calculations simultaneously reproduced the observed optical depth of CMB \citep{Planck2020} and the UV luminosity functions from galaxy observations \citep[e.g.,][]{Finkelstein2015}. In particular, they showed that the contribution from Population III stars was necessary to explain the EDGES observation \citep{Bowman2018}. It is also suggested that the star formation rate efficiency within the mini-halo should be at least ten times smaller than that of high-redshift galaxies. Thus, the 21cm global signal can provide us with statistical properties of star formation in the early Universe. Future SKA observations will be a powerful tool to constrain these statistical properties quantitatively.


\subsection{Modeling Pop III stars for semi-numerical simulations}

Not only the global signal but also the power spectrum likely reflects the properties of the Pop III stars. 
To compute the 21cm power spectra, we need to generate the spatial distribution of the signal. 
One of the most popular ways to generate the map of 21cm signal is to use semi-numerical schemes, such as 21CMFAST. 
However the original 21CMFAST does not contain a model of Pop III stars. 

One of important effects during the epoch of Pop III star formation is Lyman-Werner (LW) feedback. 
It is well known that photons with LW band easily dissociate hydrogen molecules which is the most important coolant for the Pop III star formation. 
Therefore the LW feedback regulates the star formation rate in this epoch. 

\cite{Fialkov2013} have proposed a recipe of the LW feedback for semi-numerical simulations, and \cite{Visbal2014} have succeeded in incorporating the time-dependent LW background. 
Owing to the recipe we can compute the global star formation rate density under the influence of LW feedback. 
The most of semi-numerical simulations assumed a constant escape fraction of ionizing photons for MHs hosting Pop III stars. 
Thus \cite{TT2021} have first incorporated the stellar and halo mass dependent escape fraction into 21cmFAST. The dependence is derived from spherically symmetric RHD simulations. 
They also considered the photo-heating in partially ionized gas by stellar UV radiation.

\begin{figure}
    \includegraphics[width=\columnwidth]{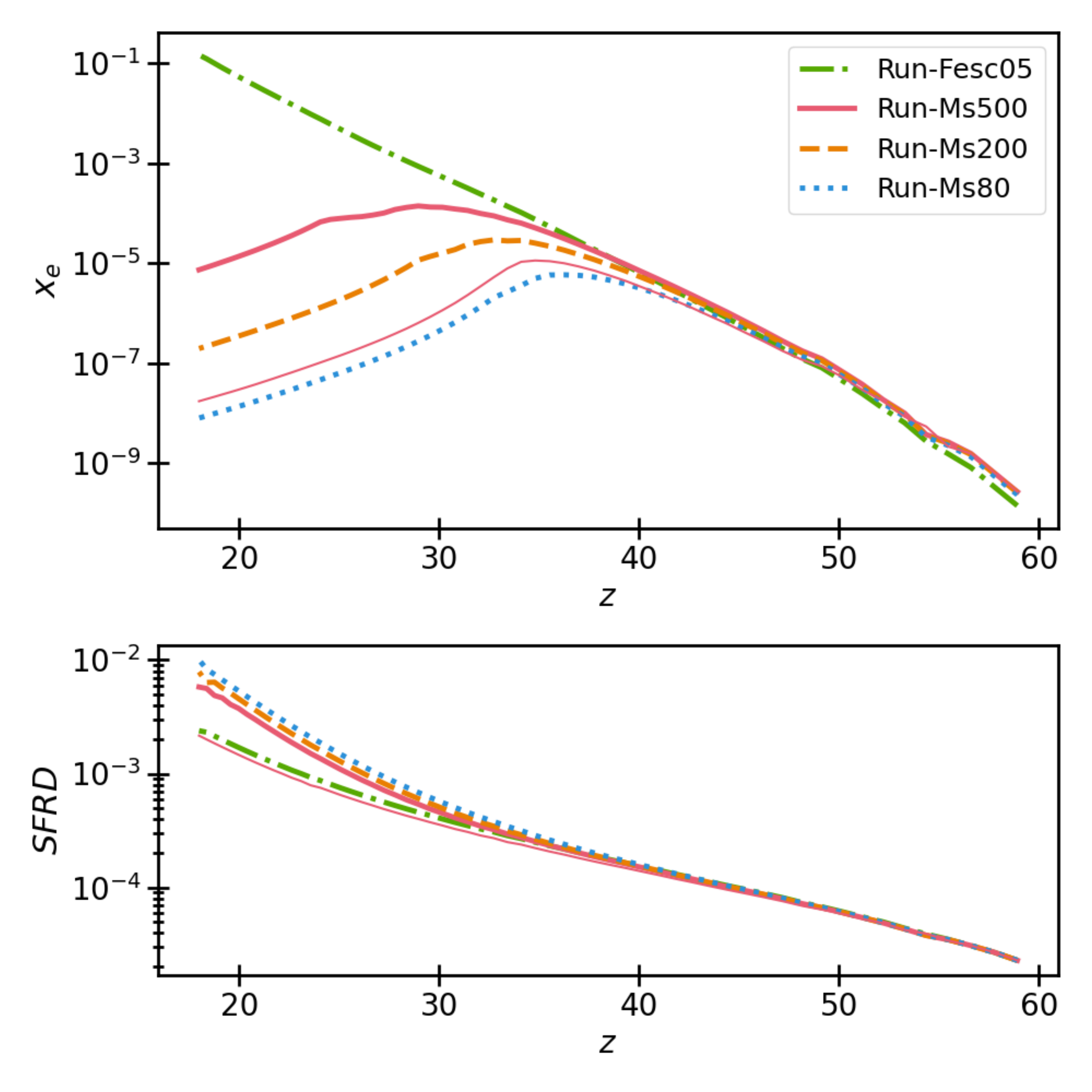}
    \caption{The box-averaged ionization fraction (top) and star formation rate density (bottom) as functions of redshift. The red solid, orange dashed, blue dotted, and the green dotted-dashed curves respectively indicate the runs of Run-Ms500, Run-Ms200, Run-Ms80, and Run-Fesc05 (taken from \cite{TT2021}, by permission of Oxford University press on behalf of the Royal Astronomical Society)). 
    }
    \label{fig:xe}
\end{figure}

Fig.\ref{fig:xe} shows the redshift evolution of ionized fractions and star formation rate densities. 
Compared to a constant escape fraction model of $f_{\rm esc}=0.5$ (named Run-Fesc05), the mass dependent escape fraction models (Run-Ms80, Run-Ms200, and Run-Ms500) show smaller ionized fraction at lower redshifts. This is because the Pop III star formation in less massive halos, which have high escape fractions, is suppressed by LW feedback. 
In addition, the ionized fraction tends to be higher with increasing stellar mass due to high escape fraction.

\begin{figure*}
    \includegraphics[width=0.8\hsize]{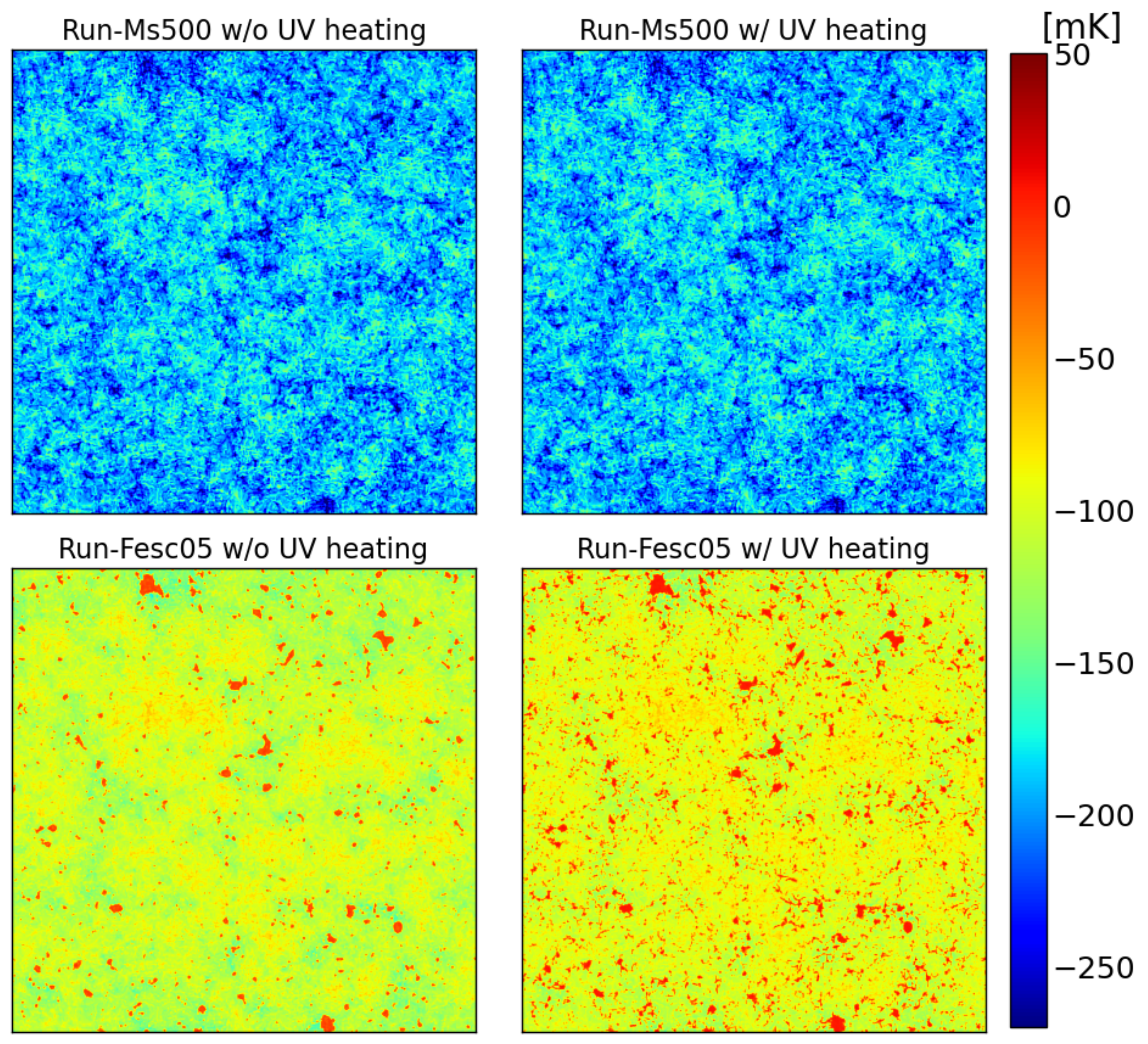}
    \caption{
    Two-dimensional maps of the 21cm differential brightness temperature at redshift 20. 
    The left and right top panels are the results from Run-Ms500 with and without UV photo-heating, respectively. 
    The left and right bottom panels are the results from  Run-Fesc05 with and without UV photo-heating, respectively. 
    Each map is 512 Mpc on a side. 
    The box-averaged ionization fractions are $x_e \sim 1.4 \times 10^{-5}$ and $\sim 5.3 \times 10^{-2}$ in Run-Ms500 and Run-Fesc05 (teken from \cite{TT2021}, by permission of Oxford University press on behalf of the Royal Astronomical Society).
    }
    \label{fig:Tbmap}
\end{figure*}

The simulated distributions of the brightness temperature are shown in Fig.\ref{fig:Tbmap}. 
In Run-Ms500, photo-heating hardly affect the brightness temperature since the ionized fraction is very small in this case. On the other hand, in the constant escape fraction model (Run-Fesc05), the overall absorption feature is relatively weak due to the high mean ionized fraction. 
Furthermore the impact of the photo-heating is remarkable on small scales.

\begin{figure}
    \includegraphics[width=\columnwidth]{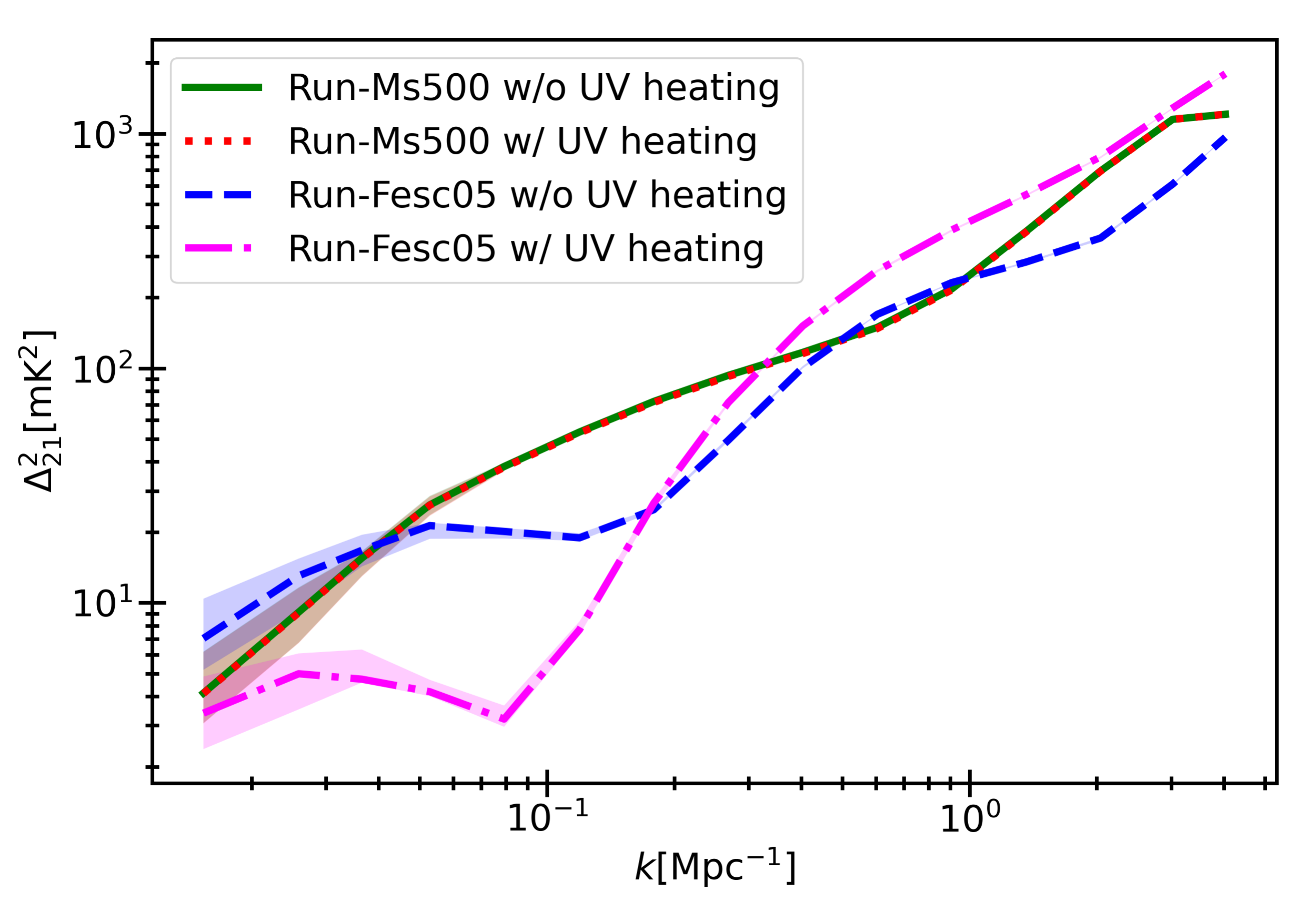}
    \caption{
    The power spectra of the 21cm differential brightness temperature at redshift 20. 
    The green solid and red dotted curves respectively indicate the results of Run-Ms500 without and with UV photo-heating. 
    The blue dashed and magenta dashed-dotted curves are these of Run-Fesc05.
    The shaded regions correspond to the 10 - 90 percentiles obtained from 10 realizations. This figure is taken from \cite{TT2021}, by permission of Oxford University press on behalf of the Royal Astronomical Society.}
    \label{fig:ps}
\end{figure}

Fig. \ref{fig:ps} show the dimensionless power spectra of the 21cm differential brightness temperature at $z=20$ in Fig.~\ref{fig:ps}. Focusing on the power spectrum obtained from Run-Ms500, we notice that the power spectrum shows relatively flat shape at the middle scale range ($k\sim 10^{-1} - 1$~[Mpc$^{-1}$]) and drops at larger scale ($k\lesssim 10^{-1}$~[Mpc$^{-1}$]). 
This trend mainly comes form the fluctuation of overdensity, and consistent with the results of \cite{Mesinger2011}. 
Since the ionization fraction is very small in Run-Ms500, the impact of the UV photo-heating on the power spectrum is not noticeable. 
On the other hand, the power spectrum in Run-Fesc05 is significantly affected by UV photo-heating. 
As shown by Fig.~\ref{fig:Tbmap}, the UV photo-heating induces fluctuations at small scales. 
In addition, the heating moderates the large-scale-inhomogeneity of the brightness temperature. 
Such relatively high ionization fraction can be achieved if the escape fraction is higher than our model and/or the star formation efficiency is higher.  

In summary, the treatment of the escape fraction in semi-numerical simulations is significantly important 
to predict the 21cm signature. 
The UV photo-heating by Pop III stars often have a notable impact on the small scale fluctuation of the 21cm signal. Therefore constructing the accurate escape fraction model is a key in this kind of study. 
In addition, it should be pointed out that the spectral shape of Pop III stars affects the efficiency of the WF coupling, and thus the 21cm brightness temperature. 
Indeed recent work by \cite{Gessey-Jones2022} shows that the WF effect starts to work from earlier epoch if the Pop III stars are typically massive. 

\section{Extracting information from the 21cm line signal}\label{sec:statistics}

\begin{figure}
    \includegraphics[width=1.0\hsize]{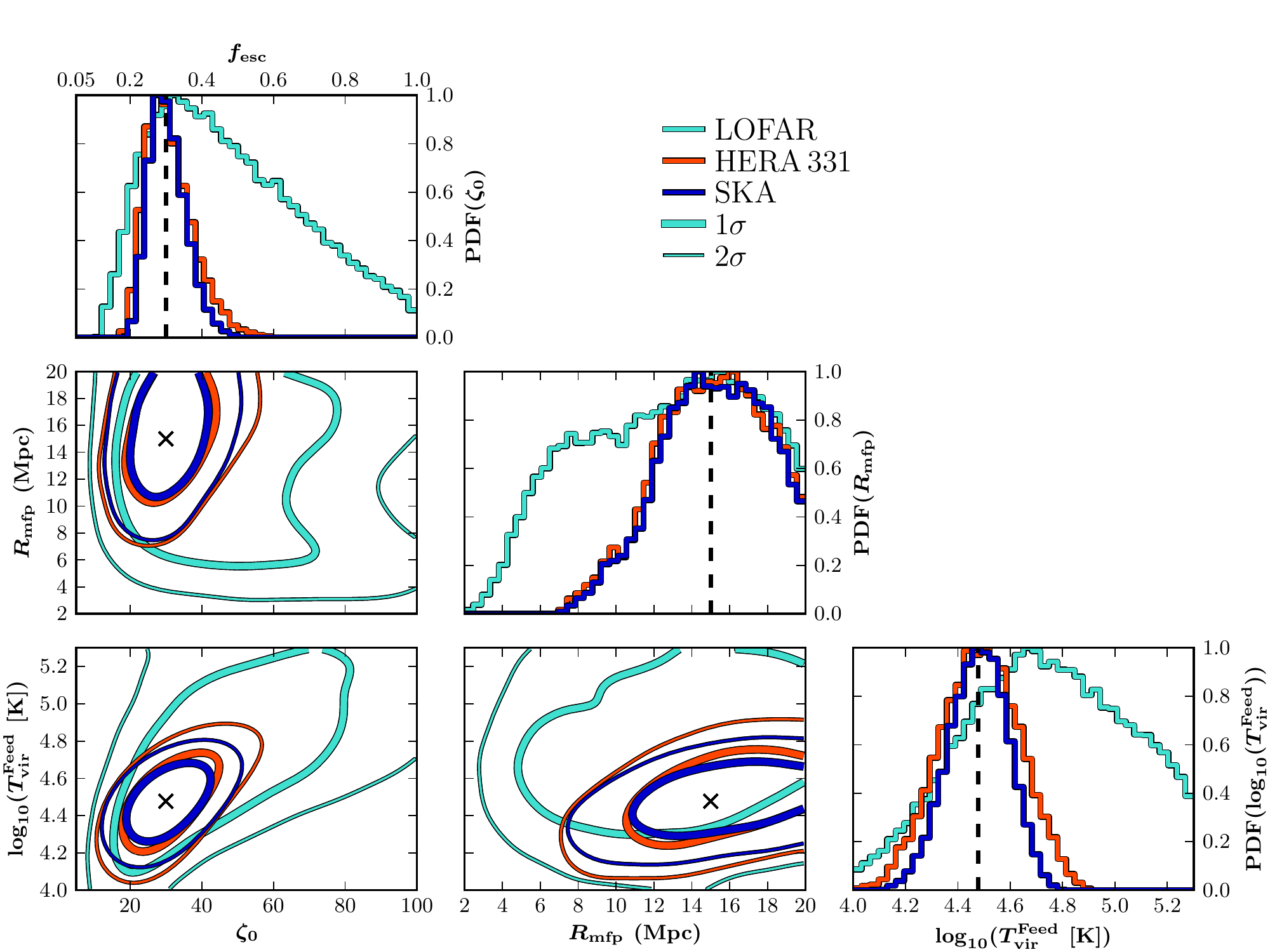}
    \caption{The contour and posterior of EoR parameters calculated by 21CMMC. In this calculation, they compare the different arrays, LOFAR, HERA 131 and SKA. This figure is reproduced from \citet{2015MNRAS.449.4246G} by permission of Oxford University press on behalf of the Royal Astronomical Society.}
    \label{fig:mcmc}
\end{figure}

So far, we reviewed the (astro)physics of the 21cm line signal. In this chapter, we see how we extract the information from the 21cm line signal. It is important to establish methods to interpret the 21cm line signal in order to extract astrophysical information. In section \ref{sec:21cmps}, we introduce the 21cm line power spectrum as a method to extract astrophysical and cosmological information. The first generation of telescopes are targeting to detect the 21cm line power spectrum and the 21cm line power spectrum brings us much information on the astrophysics during the cosmic dawn and the EoR. One of the approaches to exploit the information from the 21cm line power spectrum is to explore model parameter space. For this purpose, we often adopt Bayesian parameter inference implemented by Markov Chain Monte Carlo (MCMC). \citet{2015MNRAS.449.4246G} first developed an MCMC analysis tool called 21CMMC, which incorporates astrophysical parameters used in the 21cmFAST. As shown in Fig.\ref{fig:mcmc}, the 21CMMC estimates astrophysical parameter constraints from the 21cm power spectrum taking 21cm EoR experiments into account \citep[e.g.][]{2015MNRAS.449.4246G,2017MNRAS.472.2651G,2019MNRAS.484..933P}.

As we see above, the 21cm power spectrum with Bayesian parameter inference brings us valuable information. Actually, if the fluctuations of the 21cm line signal obey Gaussian distribution, the 21cm line power spectrum can perfectly describe the statistical properties of the 21cm fluctuations. However, we expect that the fluctuations of the 21cm line signal deviate from Gaussian distribution due to astrophysical processes. In this section, we introduce statistical approaches beyond the 21cm line power spectrum and how such statistics update our understanding. We also introduce a recent machine learning-based approach to extract information from the 21cm line signal. 

\subsection{21cm line bispectrum}\label{sec:bispectrum}

\begin{figure}
    \includegraphics[width=1.0\hsize]{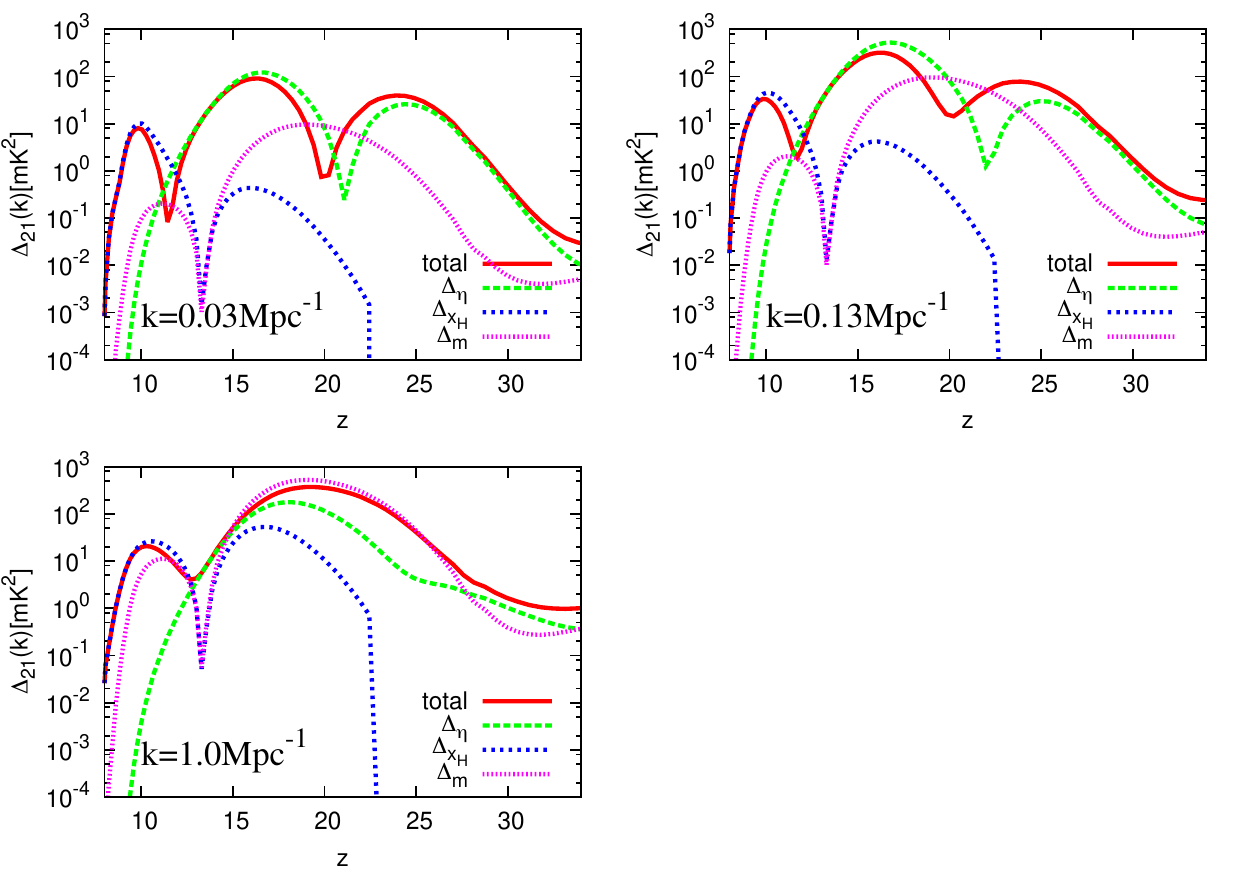}
    \caption{Total and decomposed 21cm power spectrum as functions of redshift for $k = 0.03~\mathrm{Mpc}^{-1}$ (left\ top), $0.13~\mathrm{Mpc}^{-1}$ (right\ top) and $1.0~\mathrm{Mpc}^{-1}$ (left\ bottom). Reproduced from \citet{2015MNRAS.451..467S}, by permission of Oxford University press on behalf of the Royal Astronomical Society.}
    \label{fig:21cmps_compose}
\end{figure}

At cosmic dawn, the fluctuations in the 21cm line signal are dominated by the fluctuations of the spin temperature. The fluctuation of the spin temperature is contributed from Lyman-$\alpha$ coupling and X-ray heating at cosmic dawn. During the EoR, as reionization progresses, the 21cm fluctuations are dominated by the fluctuations due to the distribution of ionized regions. In Fig.\ref{fig:21cmps_compose}, we show each components of the 21cm power spectrum. As components, we plot fluctuations of neutral fraction ($\mathrm{x_{H}}$), spin temperature ($\eta=1-T_\mathrm{S}/T_{\gamma}$), matter ($m$). In this figure, we can see what fluctuation is dominated at each redshift. The fluctuations by the spin temperature and neutral (ionized) fraction are expected to generate non-Gaussiainty in the 21cm fluctuations. To evaluate the non-Gaussianity of the 21cm fluctuations, we introduce the bispectrum for 21cm fluctuations. The 21cm bispectrum is a three-point correlation function in Fourier domain and defined by 

\begin{equation}
\langle \delta_{21}(\mathbf {k_{1}}) \delta_{21}(\mathbf{k_{2}}) \delta_{21}(\mathbf{k_{3}})\rangle =(2 \pi)^3 \delta(\mathbf{k_{1}}+\mathbf{k_{2}}+\mathbf{ k_{3}})B(\mathbf{k_{1}},\mathbf{k_{2}},\mathbf{k_{3}}),
\label{eq:bs_def}
\end{equation}
where 
\begin{eqnarray}
\delta_{21}({\bf x}) \equiv
\delta T_b({\bf x}) - \langle \delta T_b \rangle .
\end{eqnarray}
$\langle \delta T_b \rangle$ is the mean brightness temperature in the 21cm map.

To compute the 21cm line bispectrum, we need to choose three points to determine the shape of the triangle in $k$-space. To determine the shape of the triangle, we use an isosceles ansatz which is defined by $k_1=k_2=k={\alpha} k_3$ $(\alpha \ge 1/2)$. For example, the shape of the bispectrum is called "{\it squeezed type}" or "{\it local type}" in the case of $\alpha \gg 1$. In the case of $\alpha=1$ and $\alpha=1/2$, we call "{\it equilateral type}" and "{\it folded type}", respectively.

Some previous works studied the 21cm bispectrum \citep[e.g.][]{2005MNRAS.363.1049C,2007ApJ...662....1P,2015PhRvD..92h3508M}. However, these works focused on the 21cm line bispectrum as a measure of primordial non-gaussianity in matter fluctuations at the Dark ages. Thus, they did not include astrophysical effects such as the WF effect and X-ray heating. They analytically expressed the bispectrum, which is directly connected to biased matter fluctuations. 

\begin{figure}
    \includegraphics[width=1.0\hsize]{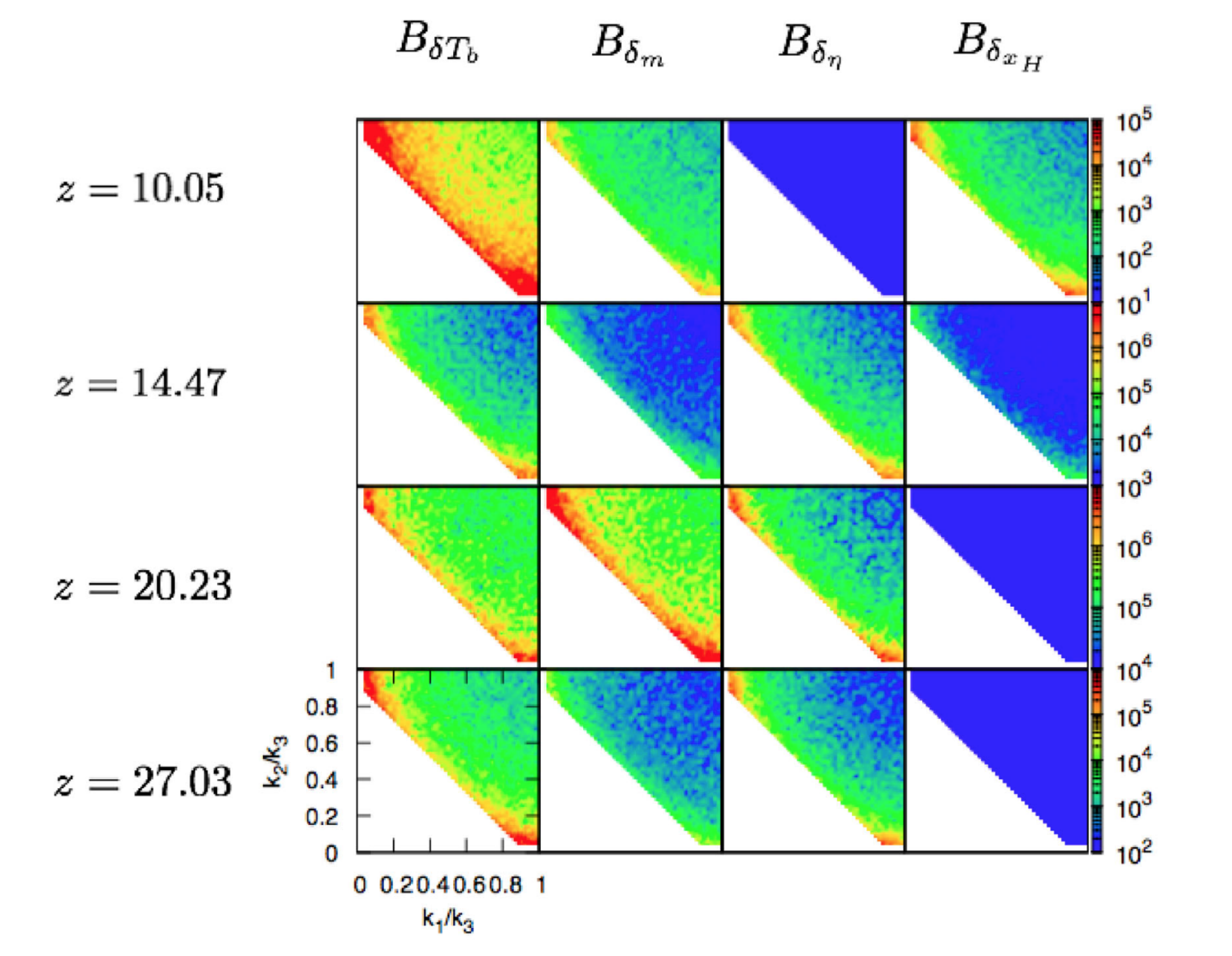}
    \caption{Contour of the 21cm bispectrum and its components in $k_1/k_3-k_2/k_3$ plane with $k_3$=1.0$\mathrm{Mpc}^{-1}$. Reproduced from \citet{2016MNRAS.458.3003S}, by permission of Oxford University press on behalf of the Royal Astronomical Society.}
    \label{fig:21cmbs}
\end{figure}

On the other hand, \citet{2016MNRAS.458.3003S} focused on the 21cm line bispectrum at the cosmic dawn and EoR. Their target is not primordial non-gauassinity, but the non-gaussianity coming from astrophysical effects. Thus, they calculated the 21cm line bispectrum directly from the 21cm image map, which includes astrophysical effects, generated by 21cmFAST \citep{2011MNRAS.411..955M}. In their work, they showed that the 21cm line bispectrum contains the information of correlation between long-wavelength and short-wavelength modes when we see the 21cm bispectrum as a function of redshift. They also showed what configuration and component are dominant in the 21cm bispectrum shown in Fig.\ref{fig:21cmbs}. This feature helps us to subtract the information from the 21cm line bispectrum. \citet{2015MNRAS.451..266Y} derived formalism to calculate the bispectrum contributed from thermal noise and evaluated the feasibility of observing 21cm bispectrum.

\begin{figure}
    \includegraphics[width=1.0\hsize]{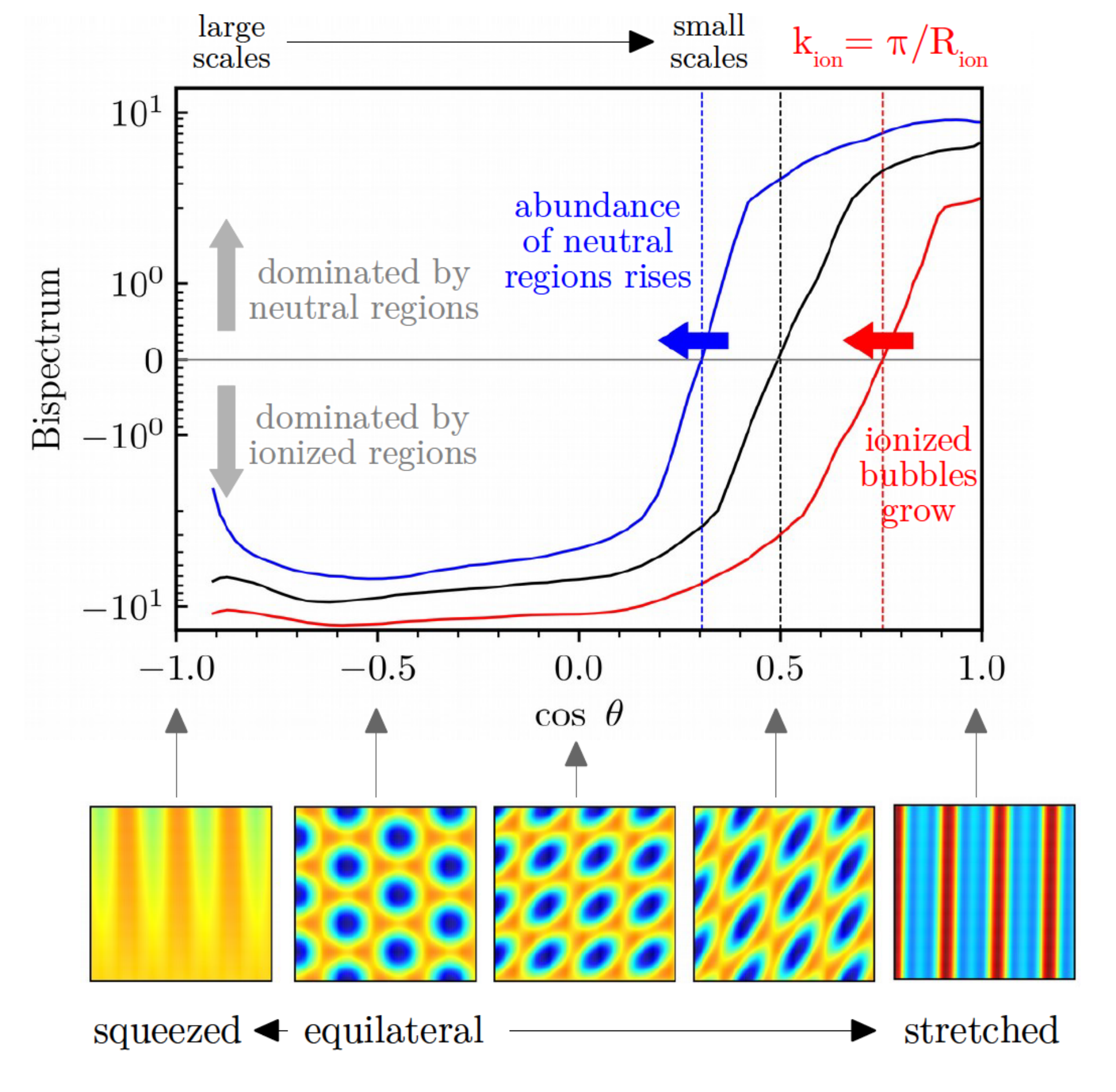}
    \caption{The 21cm bispectrum as a function of $\cos \theta$, which expresses the configuration of the bispectrum. We can see that the 21cm bispectrum has both positive and negative signs depending on whether the ionized region is dominant or the neutral region is dominant. The figure is reproduced from \citet{2020MNRAS.492..653H}  by permission of Oxford University press on behalf of Royal Astronomical Society.}
    \label{fig:sign}
\end{figure}

\citet{2017MNRAS.472.2436W} and \citet{2018MNRAS.476.4007M} have also developed an estimator to compute the 21cm line bispectrum called FFT-bispectrum estimator. Their estimator is different from \citet{2016MNRAS.458.3003S}. In the case of \citet{2016MNRAS.458.3003S}, they compute the absolute value of the bispectrum. Thus, their estimator can only evaluate the positive value of the bispectrum. Meanwhile, the FFT-bispectrum estimator uses the real part of the bispectrum, thus their bispectrum has both positive and negative signs. In \citet{2018MNRAS.476.4007M}, they showed that the negative sign of bispectrum implies that non-Gaussianity at the specific reionization is driven by size distribution and topology of the ionized regions, while the positive bispectrum comes from matter bispectrum and other various cross spectra. Furthermore, \citet{2020MNRAS.492..653H} more carefully studies how the 21cm bispectrum traces the topology of the ionized bubble. They found that the 21cm bispectrum depends strongly depends on the size distribution of ionized and neutral regions. They also found that The 21cm bispectrum changes its sign depending on whether the ionization region is dominant or the neutral region is dominant. When ionized regions are dominant, the 21cm bispectrum has a negative sign, while it has a positive sign when neutral regions are dominant (Shown in Fig.\ref{fig:sign}). This means that the position of the change of sign strongly depends on the typical size of ionized and neutral regions. As shown in eq.\ref{eq:brightness}, the brightness temperature includes the peculiar velocity of the gas, the effect of redshift space distortion (RSD) is expected to affect the 21cm bispectrum. \citet{2020MNRAS.499.5090M} studied the impact of RSD on the 21cm bispectrum. They found that  RSD affects both the sign and magnitude of the 21cm bispectrum significantly. The RSD changes the magnitude of the bispectrum by 100-200 $\%$ at the most and also flips the sign from negative to positive. Thus, they concluded that it is important to take the effect of RSD into account for correct interpretation of the 21cm bispectrum. \citet{2021MNRAS.508.3848M} evaluates the impact of light-cone effect on the bispectrum and they found that light cone effect becomes important on scales $k_1 \lesssim 0.1 \mathrm{Mpc^{-1}}$.

While the works introduced above consider the 21cm line bispectrum at the EoR to study non-Gaussianity driven by ionized regions, there are some works which focuses on the bispectrum at cosmic dawn \citep[e.g.][]{2021MNRAS.502.3800K,2021arXiv210808201K}.  For example, \citet{2021MNRAS.502.3800K} studied the impact of both RSD and spin temperature on the 21cm bispectrum during cosmic dawn. They found that the effect of spin temperature impacts on the magnitude of the bispectrum for small triangle configuration when Lyman-$\alpha$ coupling is saturated. They also found that RSD affects the magnitude of the bispectrum and changes the sign of the bispectrum in the case of a large triangle configuration.

The 21cm bispectrum is also applied for parameter estimation as similar as power spectrum \citep[e.g.][]{2017MNRAS.468.1542S,2021arXiv210202310W,2021arXiv210807279T}. In \citet{2017MNRAS.468.1542S}, they performed Fisher forecast for EoR parameters with the 21cm line bispectrum. They found that the bispectrum puts tighter constraints on the parameters than the power spectrum and the combination of bispectrum and power spectrum gives more tighter constraints. \citet{2021arXiv210202310W} and \citet{2021arXiv210807279T} also studied how the 21cm bispectrum impacts on parameter constraints. They performed Bayesian parameter inference implemented by MCMC with the 21cm line bispectrum and found that 21cm bispectrum improves the constraints on parameters.

\subsection{One-point statistics}
\label{sec:moment}

In previous sections, we introduced the 21cm power spectrum and bispectrum as statistical quantities. The power spectrum and bispectrum are quantities in Fourier space and we also can statistical quantities in real space. Here, we introduce one-point statistics. As a simple case of one-point statistics, some works studied probability distribution function (PDF) of 21cm fluctuations\citep[e.g.][]{2008MNRAS.384.1069B,2010MNRAS.406.2521I,2010MNRAS.408.2373G}.  \citet{2008MNRAS.384.1069B} presented and studied a PDF of the difference between two points of 21cm brightness temperature. They showed that the PDF can measure statistics that directly depend only on ionized distribution while the usual correlation function is determined by a complicated mixture of density field and ionization field. Some works considered variance and skewness of PDF of 21cm fluctuations, which are higher-order moments of PDF \citep[e.g.][]{2014MNRAS.443.3090W,2015MNRAS.454.1416W,2015MNRAS.451..467S,2021A&A...653A..58G}. The variance is the expectation of the squared deviation of a random variable from its mean value and the skewness is a measure of the asymmetry of PDF of a variable. The variance and skewness of the 21cm map are calculated by

\begin{eqnarray}
&& \sigma^2 = \frac{1}{\delta T_b} \sum_{i=1}^{N} \big[ \delta T_b - \langle \delta T_b \rangle \big]^2
\label{eq:variance} \\
&& \gamma = \frac{1}{N \sigma^3} \sum_{i=1}^{N} \big[ \delta T_b - \langle \delta T_b \rangle \big]^3,
\label{eq:skewness}
\end{eqnarray}
where $N$ is the number of pixels of the maps and $\langle \delta T_b \rangle$ is average value of $\delta T_b$ in all pixels. The variance (skewness) is associated with power spectrum (bispectrum) as follow (see \citet{2016PASJ...68...61K}.

\begin{equation}
\sigma^{2}=\langle \delta T_b\rangle^{2} \int \frac{d^{3} k}{(2 \pi)^{3}} P(\mathbf{k})
\end{equation}
\begin{equation}
\gamma=\langle \delta T_b\rangle^{3} \int \frac{\mathrm{d}^{3} k_{1}}{(2 \pi)^{3}} \int \frac{\mathrm{d}^{3} k_{2}}{(2 \pi)^{3}} B\left(\boldsymbol{k}_{1}, \boldsymbol{k}_{2},-\boldsymbol{k}_{1}-\boldsymbol{k}_{2}\right)
\end{equation}

 \citet{2014MNRAS.443.3090W} applied one-point statistics to distinguish models of reionization. They calculate variance $\sigma^2$ and skewness $\gamma$ for the 21cm brightness temperature as one-point statistics. They consider the following 4 reionization models. Reionization is driven by 1.large ionized bubbles around over-dense region(global inside-out), 2.small ionized bubbles in over-dense region (local inside-out), 3.large ionized bubbles around under-dense regions (global outside-in) and 4.small ionized regions around under-dense regions (local outside-in).They found that negative skewness is found only in the global inside-out model. They also found that one-point statistics enable us to distinguish them even by pre-SKA experiments although it is difficult to distinguish models using 21cm line power spectrum in pre-SKA experiments. While \citet{2014MNRAS.443.3090W} consider the one-point statistics at EoR, some works consider the impact of X-ray heating on one-point statistics of 21cm brightness temperature at cosmic dawn\citep[e.g.][]{2015MNRAS.454.1416W,2017MNRAS.468.3785R}. For example, \citet{2015MNRAS.454.1416W} found that the peaks of the skewness and variance of 21cm brightness temperature in the redshift evolution are sensitive to X-ray efficiency. They also found that the amplitude of variance is sensitive to the hardness of X-ray SED.
 
\citet{2015MNRAS.451..467S} applied one-point statistics to give a physical interpretation of the 21cm line power spectrum at cosmic dawn. They deeply studied the dip and peaks that appeared in the 21cm power spectrum as a function of redshift (see bottom of Fig.\ref{fig:21cmps}) with one-point statistics. They found that the redshift evolution of dip strongly depends on X-ray heating and the skewness of 21cm brightness temperature becomes a good indicator when X-ray heating becomes effective.

Although the works shown above include thermal noise, they do not take instrumental effects into account. Some works consider more realistic situation \citep[e.g.][]{2009MNRAS.393.1449H,2014MNRAS.443.1113P,2018MNRAS.474.4487K}. \citet{2009MNRAS.393.1449H} suggested one-point statistics as a probe to characterize cosmic 21cm signal after cleaning of foregrounds, thermal noise, and instrumental effects. They performed simulations of cosmological 21cm signal, foregrounds, and instrumental noise and make realistic mock data cube. They fit foregrounds with a three-order polynomial in log frequency to each pixel. After the fitting, they compute skewness from residuals and they found that they can recover main features (dip appeared at the beginning of reionization and rise that appears as reionization proceeds) of the redshift evolution of skewness in the cosmological 21cm signal.\citet{2009MNRAS.393.1449H} showed the skewness is useful to extract the information of cosmological 21cm signal from realistic 21cm data. \citet{2014MNRAS.443.1113P} investigates the extraction of variance of 21cm line signal and constrains global history of the EoR assuming LOFAR experiment. They showed that the LOFAR with 600 hours of integration time can detect variance of 21cm signal and recover parameters of the global evolution of the 21cm line signal. Similar to the LOFAR experiment, \citet{2018MNRAS.474.4487K} showed HERA experiment can also detect characteristic features of one-point statistics.

Another application of one-point statistics is parameter constraints. \citet{2016PASJ...68...61K} performed Fisher analysis for EoR parameters with one-point statistics assuming LOFAR and MWA experiments. They showed that the combination of the variance and skewness can strongly constrain the EoR parameters.

As described above, one-point statistics is very useful to distinguish the EoR models, to characterize 21cm signal from realistic mock data including the cosmic 21cm signal, the effects of foregrounds, and experimental noise, to give a physical interpretation of the behavior of the 21cm line power spectrum, and to constrain model parameters.

\subsection{Machine learning approach}\label{sec:ML}

\begin{figure}
\includegraphics[width=1.0\hsize]{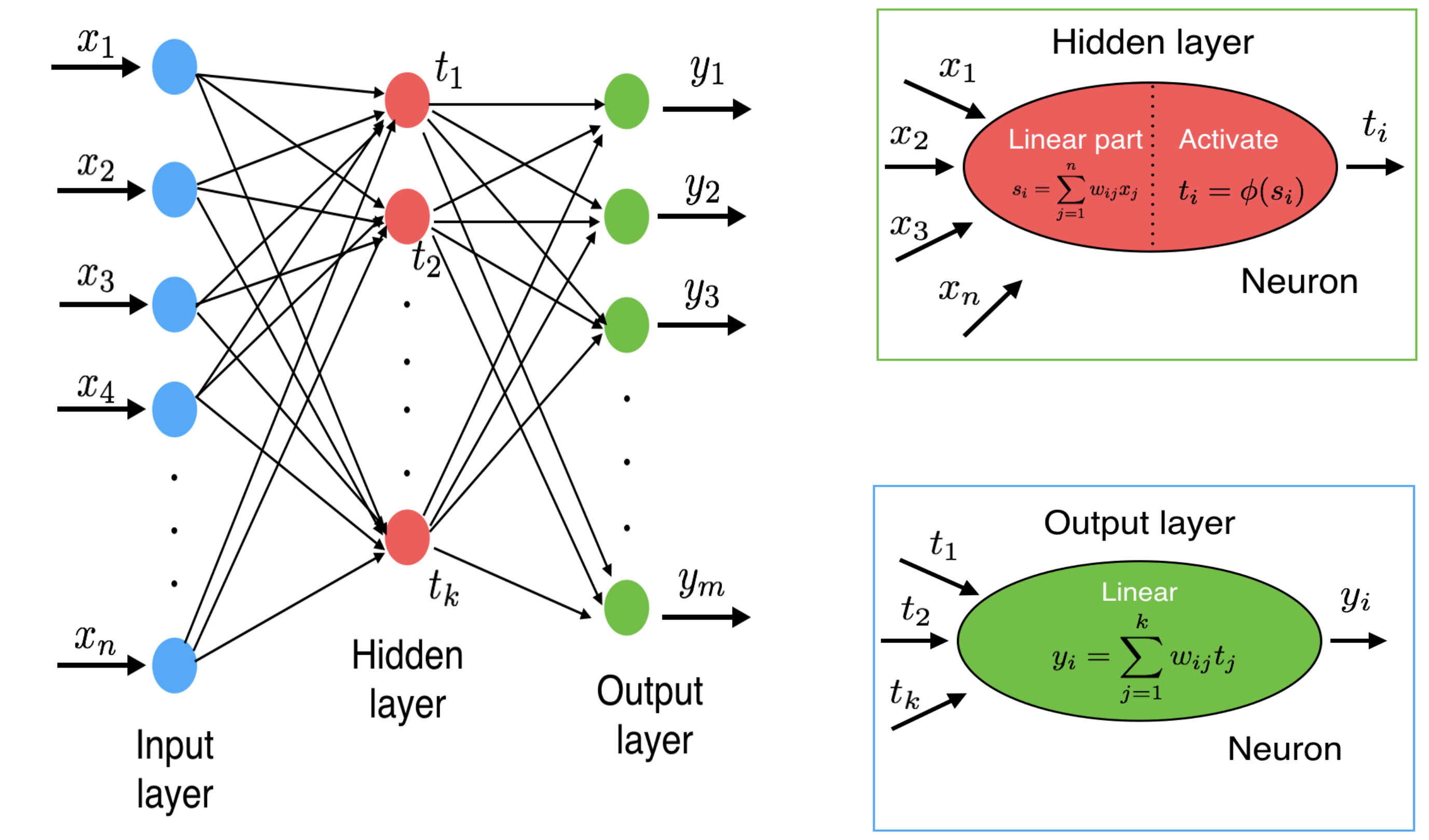}
\caption{Typical architecture of an artificial neural network.  The architecture of the ANN consists of an input layer, a hidden and an output layer of neurons. Each neuron connects the neurons in the next layer. The figure is reproduced from \citet{2017MNRAS.468.3869S}  by permission of Oxford University press on behalf of the Royal Astronomical Society.}
\label{fig:fig1}
\end{figure}

So far, we have discussed the analysis of the 21cm line signal using conventionally used methods. In this section, we introduce an approach based on machine learning methods that are widely used these days.  The concept of the machine learning approach is that the machine itself automatically improves itself by learning from data. In particular, Artificial Neural Networks (ANNs) are often used in the context of 21cm study. The ANNs are mathematical models inspired by the neuron network in our brain. The main purpose of ANNs is to construct approximate functions which associate input with output by using datasets. This process is called training. To construct such an approximate function, the ANN has to learn from training data. The architecture of the ANN consists of three layers: the input layer, the hidden layer, and the output layer.  Each of them has several neurons and each neuron is connected to other neurons as shown in Fig.\ref{fig:fig1}. After training, the trained architecture can be applied to unknown data called test data and predict output values from input values.

\begin{figure}
    \includegraphics[width=1.0\hsize]{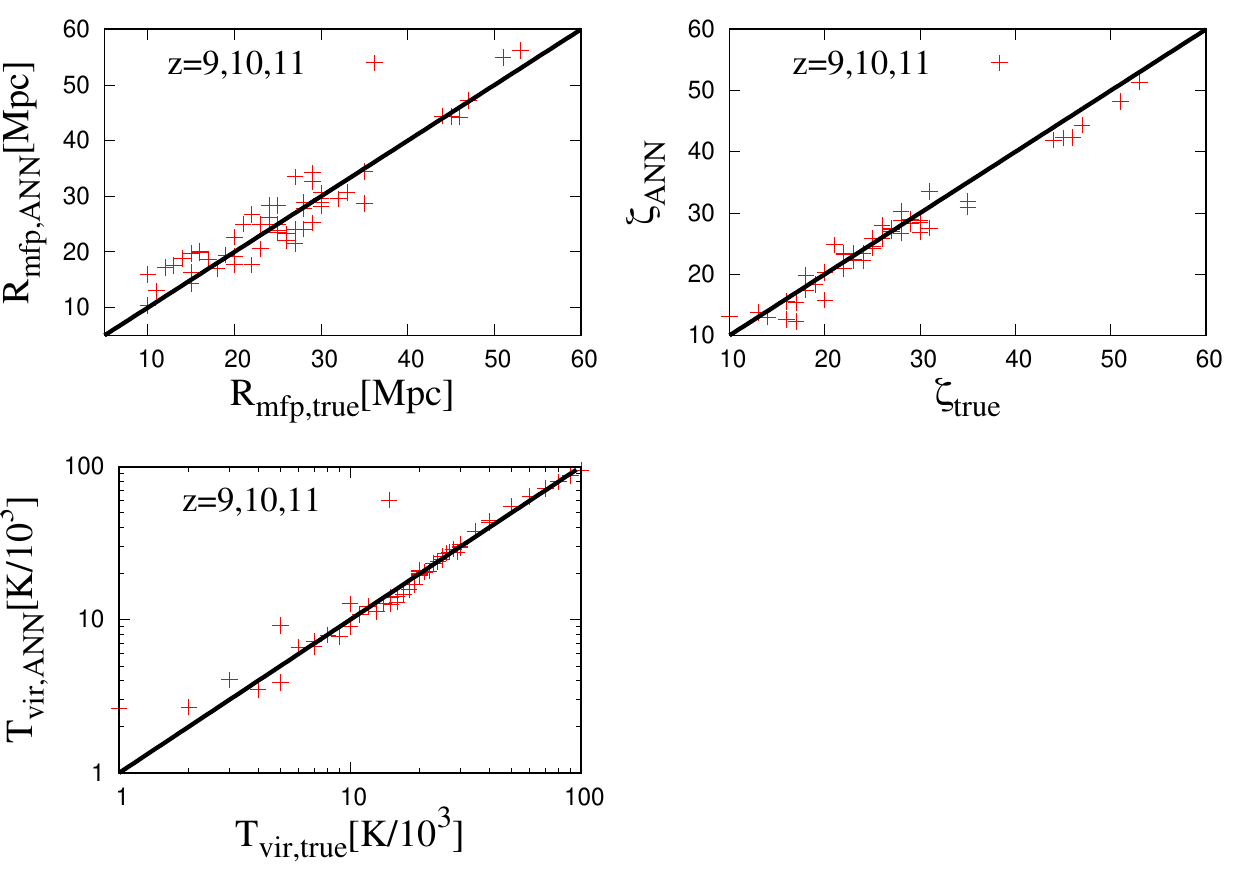}
    \caption{The comparison of the EoR parameters obtained from the 21cm line spectrum with the ANN against true values. They use the 21cm line power spectrum at $z$=9,10,11 and include both thermal noise and sample variance. The figure is reproduced from \citet{2017MNRAS.468.3869S} by permission of Oxford University press on behalf of the Royal Astronomical Society.}
    \label{fig:ann_para}
\end{figure}

In the context of the 21cm study, \citet{2017MNRAS.468.3869S} first introduced the ANN for parameter estimation. In \citet{2017MNRAS.468.3869S}, the architecture of the ANN is constructed by the 21cm line power spectrum and EoR parameters. They trained the architecture by training data which consists of the 21cm line power spectrum and EoR parameter. With the trained architecture, they predict EoR parameters from the 21cm line power spectrum shown in Fig.\ref{fig:ann_para}. They showed that the ANN successfully recover EoR parameters from the 21cm line power spectrum with high accuracy compared to the true values.

In \citet{2017MNRAS.468.3869S}, they recovered the EoR parameter used in the 21cmFAST from the 21cm line power spectrum with the ANN. In their work, although they take thermal noise into account, they do not include the effect of the foreground. \citet{2021arXiv211213866C} evaluates the performance of the ANN in the case that they include foreground. They showed that the ANN can recover EoR parameters with 81-90\% of accuracy even if they include foreground in the case of the SKA experiment. In the context of parameter recovery with the ANN, \citet{2020MNRAS.491.4031C} used the ANN to extract the 21cm global signal parameters from mock simulation data which includes the effects of foreground, instrumental effects, and noise and \citet{2021MNRAS.502.2815C} applied the ANN to predict signal parameters from EDGES data as inputs.

The works shown above applied the ANN to recover parameters from 1-dimension inputs such as power spectrum and global signal. On the other hand, some works have adopted a deep neural network (or deep learning) which has multiple hidden layers to treat image maps. One of the most often used deep neural networks is Convolution Neural Networks (CNN). Some works developed a method to extract cosmological and astrophysical information from 21cm image maps or light cone maps. \citep[e.g.][]{2019MNRAS.484..282G,2020MNRAS.494.5761H,2021arXiv210503344Z,2021PASP..133d4001B,2020JKPS...77...49K,2020MNRAS.493.5913L,2022MNRAS.509.3852P,Zhao_2022}. The advantage of applying the CNN is that they can exploit more information than 1-dimension inputs because image data is regarded as multi-dimension vector and it contains more information than 1-dimension values such as 21cm line power spectrum and global signal. Actually, these works have shown astrophysical and cosmological parameters can be determined with high accuracy with deep neural networks.

Another application of the ANN is an emulator. The purpose of the emulator is to build ANN architecture trained by prepared simulation data in order to quickly calculate output values for input values. In the context of 21cm studies, the emulator is used to compute statistical quantities(power spectrum and bispectrum) and global signals for given input astrophysical parameters. The emulator has an advantage when we combine it with MCMC. To perform MCMC, it is necessary to run simulations each time for different parameter samplings to compute the likelihood function for comparing observation data (or mock data) with theoretical models. However, the emulator can save calculation costs for running simulations because the emulator returns output value for given input parameters. Recent works have developed emulators for the calculation of the 21cm power spectrum and global signal for given input astrophysical parameters \citep[e.g.][]{2017ApJ...848...23K,2018MNRAS.475.1213S,2020MNRAS.495.4845C,2021MNRAS.508.2923B,2021arXiv210705581H,2022arXiv220108205S}. For example, \citet{2018MNRAS.475.1213S} constructed the emulator to calculate the 21cm line power spectrum from given input EoR parameters. They showed that the emulator can speed up the parameter sampling speed by 3 orders of magnitudes when performing MCMC.

As another example of the application of the ANN, \citet{2022RAA....22c5027S} applied the ANN to recover ionized bubble size distribution from the 21cm line power spectrum. They showed the ANN can successfully recover ionize bubble size distribution from 21cm line power spectrum by using ANN with a few $\%$ accuracy. The philosophy of their work is to recover statistical quantity from another statistical quantity with ANN. With the same philosophy, \citet{2021MNRAS.506..357Y} applied Generative Adversarial Networks (GAN) to generate 21cm image maps from the distribution of Lyman-$\alpha$ galaxies. The GAN is often used to generate new data with the same statistics as training datasets. They showed that the distribution of 21cm brightness temperature can be reproduced from Lyman-$\alpha$ galaxies distribution with correlation function of $\sim 0.5$ at $k < 0.1 \mathrm{Mpc}^{-1}$.

\section{Synergy with other experiments}\label{sec:21cmCC}
Ideally, we want to use 21cm line signal itself to extract astrophysical information by using approaches introduced in section \ref{sec:statistics}. However,we face some difficulties with 21cm line observation.
For example, current upper limits on the 21cm power spectrum are not limited by the thermal noise but enormous foreground contamination and systematics. The 21~cm line is more than 4 orders of magnitude fainter than the foregrounds such as synchrotron emission from galactic and extragalactic radio sources \citep[e.g.][]{ChapmanJelic2019}. Thus, the detection of the 21~cm line is impossible without foreground mitigation techniques. The foregrounds are expected to be spectrally smooth in frequency space, and the foreground power effectively decreases at small scale in Fourier modes of $k_{\parallel}$ which corresponds to spacial scales along the line of sight. The mode mixing effect due to interfermeteric nature generates spectrally non-smooth contamination and propagates the foreground contamination to higher $k_{\parallel}$ at higher $k_{\perp}$, which is angular scales perpendicular to the line of sight. Consequently there are Fourier spaces called the ``EoR window'' (e.g. \cite{2014PhRvD..90b3019L}) where the 21~cm line signal can dominant the foreground. Such spectrally well behaved foregrounds also can be removed using foreground removal methods \citep{Chapman2015aSKA} such as FastICA \citep[][]{Chapman2012MNRAS}, GMCA \citep[][]{Chapman2013MNRAS}, and GPR \citep[see e.g. ][]{Mertens2018MNRAS;GPR,Kern2021MNRAS}. Theoretical works have shown that these strategies are promising to reduce the foreground contamination. However, these methods can fail to perform ideal performance due to spectrally non-smoothness of foregrounds caused from mode-mixing, instrumental systematics and data analysis \citep{Barry2016CalibrationDesigned2}.

In order to overcome the systematics and reduce the foreground contamination, the 21~cm line cross correlation (CC) with other observable should be useful. The low-$z$ 21~cm line has been detected by combining the foreground removal and the CC with the galaxies \citep[e.g.][]{2010Natur.466..463C}. Thus, the CC is the promising strategy to detect the 21~cm line from the EoR. The CC is also important to validate the detection of 21~cm auto power spectrum since the auto power spectrum is easily biased by any systematic errors. The CC itself should have unique information of various astrophysics at high-$z$. Therefore the CC is one of primary sciences for the SKA \citep{2015aska.confE...4C,2015aska.confE...8J}. Previous works have suggested several possible partners to the CC. In this section, we introduce possible CCs with the 21~cm line before the EoR by referencing  \cite{2015aska.confE...4C,2015aska.confE...8J,2019BAAS...51c.282C}.

\subsection{Cosmic Microwave Background, Near Infrared Background And X-ray Background}

Cosmic Microwave Background (CMB) is one of the most powerful proves of reionization. For example, the optical depth to the Thomson scattering for the CMB photon depends on the total amount of ionized fraction from the last scattering surface to the present, and the recent CMB observations constrain the reionization history \citep{2020A&A...641A...6P}. Furthermore, the CMB photon is scattered by dense ionized bubbles during the EoR, and the CMB should obtain fluctuations due to the peculiar motion of the ionized bubbles. This is the kinematic Sunyaev-Zel'dovich (kSZ) effect which generates fluctuations correlating with the 21~cm line. Many works have predicted the cross correlation signal analytically and numerically \citep[e.g.][]{2004PhRvD..70f3509C,2005MNRAS.360.1063S,2006ApJ...647..840A,2008MNRAS.384..291A,2010MNRAS.402.2279J,Alvarez2016A}. The signal is quite useful statistics because the amplitude and sign of the signal depend on the ionized history and size of ionized bubbles. However, the correlation is not strong enough, the detection would be tough due to the cosmic variance of the primary fluctuation of the CMB\citep[e.g.][]{2010MNRAS.402.2279J,Alvarez2016A} and foregrounds \citep{2019MNRAS.483.2697Y}. 

The CMB E-mode and B-mode are also useful partner for CC \citep{Tashiro2008MNRAST,Kadota2019PhRvD}. The CC can give constraints on the evolution of the reionization, while the detectability is not high due to lower correlation coefficient. In \cite{Tashiro2010MNRAS}, they have investigated the detectability of the 21~cm CC with the anisotropy in the CMB temperature and polarization using an analytic calculation, and found that the CPS with the temperature is detectable with the Planck and the SKA, while the CMB polarization is not. 

Alternatively, feasibility for the 21cm-$\rm kSZ^2$ CC has been investigated in \cite{Ma2018MNRAS}. The signal can be measured by filtering out the primary CMB fluctuation using Wiener filter \citep{2004ApJ...606...46D}. Recently the 21cm-kSZ-kSZ bispectrum has been proposed as well \citep{2020ApJ...899...40L}. On the other hand, future observation of CMB polarization with CMB-S4 \citep{2019arXiv190704473A}, PICO \citep{2019arXiv190210541H} could reconstruct the 2D map of optical depth to Thomson scattering $\tau_{\rm eff}$ \citep{2009PhRvD..79d3003D}. Since the $\tau_{\rm eff}$ depends on the fluctuation of ionized fraction and fluctuation of gas density at each line of sight, the 2D $\tau_{\rm eff}$ also correlates with the 21~cm signal \citep{Anirban2020JCAP}. Noting that although both kSZ and $\tau_{\rm eff}$ maps are provided as 2D images, we can expect unique correlation with the 21~cm signal at each redshift. This enables us extract the redshift evolution of ionized fraction.

Near Infrared Background (NIRB) is also a possible partner for the 21~cm CC \citep{Wyithe2007MNRAS, Fernandez2014MNRAS,2014ApJ...790..148M}. The correlation is generated via the highly redshifted UV photon emitted from stars and black holes \citep[e.g.][]{Kashlinsky2005PhR,Kashlinsky2018RvMP} which can be the source of excess component in NIRB. As shown in these previous works, the correlation coefficient expected to be negative after $\approx$ 50\% of H{\sc i} is ionized since the star ionizes neutral hydrogen. In \cite{2014ApJ...790..148M}, they have shown that the SKA1 Low can measure the CPS by combing CIBER-2 survey \citep{2014SPIE.9143E..3NL}.

The analysis of unresolved X-ray background has indicated some contribution from high-$z$ sources \cite[e.g.][]{Cappelluti2012}. While the high-$z$ component might be less than a few percent \citep[][]{Ma2018MNRASXRB}, the 
21~cm line and X-ray background (XRB) CC can be a useful tool to reveal the XRB source as suggested in \cite{2009RAA.....9...73S,2016RAA....16..132L,Ma2018MNRASXRB}. In \cite{Ma2018MNRASXRB}, they studied the CC using radiative transfer simulation \citep{2020MNRAS.498.6083E}. The CC signal shows positive correlation at the Epoch of Heating and negative correlation at the EoR. In assumption of SKA1 Low, the detectability is limited by the noise of X-ray observation, and thus the CC might require a mature of future X-ray survey.

The CC between XRB and cosmic infrared background (CIRB) has suggested presence of high-$z$ direct collapse black holes \citep{2018ApJ...864..141L,2016ApJ...832..104M,2013ApJ...769...68C,2019ApJ...883...64L} as a source of excess fluctuation in CIRB. The source is still controversial, and the CC with the 21~cm line and these background map might be an evidence for the high-$z$ components.

\subsection{Line Intensity Mapping And High-$z$ Galaxy}

Synergy with other line intensity mapping also has a potential to detect the 21~cm line without foreground contamination \citep{2019BAAS...51c.282C}. For example, the CO(1-0) molecule line at 115~GHz in rest from its rotational transitions is a tracer of molecular gas and star formation at high-$z$. While the CO model is uncertain at high-$z$, the CC with the 21~cm can prove the models \citep{Gong2011_21cmCO,Lidz2011_21cmCO,2020arXiv200902766Z}. The highest band of SKA1 Mid covers CO line from $z>7.3$. Thus, the 21cm-CO CC is a good synergy for the SKA itself.

As a tracer of star formation, intensity mapping of [CII] fine-structure emission line can be useful while there are large uncertainty at high-$z$ \citep[e.g.][]{2019MNRAS.490.1928Y,2022arXiv221208077P}. The 21~cm line should have anti-correlation with [CII] intensity map at large scales \citep{2012ApJ...745...49G,2015ApJ...806..209S,2019MNRAS.485.3486D}. The CC with the 21~cm line can constrain the model of [$\rm CII$] to star formation rate and reionization. For example, \cite{2019MNRAS.485.3486D} have shown that 21cm-[CII] CC can be detected by combing LOFAR/SKA1Low and current/future [CII] survey \citep[CONCERTO][]{2018IAUS..333..228L}.

High-$z$ [OIII] line emitting galaxy has been observed by ALMA \citep{2016Sci...352.1559I,2017A&A...605A..42C,2017ApJ...837L..21L,2019PASJ...71..109H,2019PASJ...71...71H,2020ApJ...896...93H}. In \cite{2019MNRAS.tmp.2236M}, they have predicted the CC of 21cm-OIII galaxy and OIII intensity mapping using numerical simulation where line luminosity of [OIII] is computed as in \cite{Moriwaki2018MNRAS}. They have shown that the sign of the CPS at $\rm k=0.1~Mpc^{-1}$ is positive at $z\approx 10$ due to hotter H{\sc i} gas in high dense regions and becomes negative at $z<8$ due to large ionized bubbles around the galaxies. The SKA1 Low can measure the CC in tandem with a large survey of [OIII] line emitting galaxy.

The Lyman-$\alpha$ is emitted from galaxies and IGM mainly due to recombination, and the radiation can be measured as Lyman-$\alpha$ intensity mapping. The CC between the 21~cm line and Lyman $\alpha$ intensity map can be detected. While the contamination dominated by the galactic terms, by combining H-$\alpha$ line, the IGM contribution will be separated \citep{2013ApJ...763..132S,2017ApJ...849...50N,2017ApJ...848...52H,2022MNRAS.tmp..487C,2021MNRAS.506.1573H}. 


The CC with 21~cm line and galaxies has succeeded in low-$z$ H{\sc i} intensity mapping \citep{2010Natur.466..463C,2013ApJ...763L..20M,2017MNRAS.464.4938W,2018MNRAS.476.3382A}. Therefore, high-$z$ galaxy is one of the most promising partner for the EoR 21~cm line CC investigated initially in \cite{2007ApJ...660.1030F,2007MNRAS.375.1034W}. Typically, as the galaxies create ionized bubbles, the 21~cm line dims around the observed galaxies. Thus, the CC between the 21~cm line and the number density of galaxies is expected to be negative. For the CC with the 21~cm line at the EoR, many recognized the possibility of CC with the Lyman-$\alpha$ emitting galaxies which have already been observed using narrow band filter \citep[][]{2018PASJ...70S..13O}.

One of promising partner is the high-$z$ Lyman-$\alpha$ emitters (LAEs). Previous studies has investigated the CC theoretically \citep[e.g.][]{2009ApJ...690..252L,2016MNRAS.459.2741S,2017ApJ...846...21F,2018MNRAS.479L.129H,2020MNRAS.496..581H}. The detectability of the 21cm-LAE CC has been explored in assumption of the MWA observation \citep[][]{Park2014}, LOFAR \citep{2013MNRAS.432.2615W,Vrbanec2016}, SKA1 Low \citep{2017ApJ...836..176H, Kubota2018MNRAS.479.2754K,Weinberger2020} in tandem with the HSC/PFS LAE observation. Recently, in \cite{2019BAAS...51c..57H,2020MNRAS.492.4952V}, they have explored that the SKA1 Low has an opportunity to collaborate with the Roman Space Telescope \footnote{{http://roman.gsfc.nasa.gov}} in the context of 21cm-LAE correlation.
For example, in series papers \citep{Kubota2018MNRAS.479.2754K,Yoshiura2018MNRAS.479.2767Y,Kubota2020MNRAS.494.3131K}, they investigated detectability of the 21cm-LAE CPS by employing a radiative transfer (RT) reionization simulation \citep{2016arXiv160301961H} and mock LAE samples. As Subaru HSC surveys have produced massive LAE catalogue in large field of view $\sim$ 20~$\rm deg^2$, they have focused on the 21cm-LAE CPS analysis in assumption of future SKA1 Low and HSC with followup spectrogragh observation using the Prime Focus Spectrograph\footnote{https://pfs.ipmu.jp} \citep{2014PASJ...66R...1T}. In \cite{Kubota2018MNRAS.479.2754K}, they showed the detectability of the 21cm-LAE CPS with assumption of the MWA/SKA1 Low and HSC/HSC-PFS observation. For the 21~cm line observation by the MWA and SKA1 Low, we assumed 1000 hours of observation time. The error is dominated by the sample variance and therefore only increasing the field of view can reduce the error. In \cite{Kubota2020MNRAS.494.3131K}m, they showed 21cm-LAE CPS can constrain the model of LAEs. In \cite{2018PASJ...70...55I} they have introduced stochastic Lyman-$\alpha$ production and transmission following the dispersing escape fraction as shown in \cite{2018MNRAS.477.5406Y}. While HSC have published LAE catalogue at $z=5.7$, they have focused only on the LAE at $z=6.6$ and $7.3$ because the reionization is expected to end by the redshift. However, recently, the Lyman-$\alpha$ fluctuation indicates the delay end of reionization, and the CPS at $z=5.7$ can verify the delay reionization model \citep{Weinberger2020}.

The foreground of the 21~cm line observation has no correlation with the distribution of the high-$z$ galaxy, and thus the foreground does not bias the power of the CPS. On the other hand, the statistical variance can create large error which is typically ignored in previous works. For example, in \cite{Yoshiura2018MNRAS.479.2767Y}, they showed that the statistical error dominated by the foreground contamination must be reduced by the foreground removal and avoidance. This is not only for the cross correlation with high-$z$ galaxies but also for any cross correlation analysis. We note that the error of foreground can be reduced not only foreground removal but also increasing the survey area.

We here demonstrate the detectability of 21cm-LAE CPS by assuming the SKA1 Low and HSC-PFS observations at redshift $z=6.6$. We employ the RT simulation and LAE models used in \cite{Kubota2018MNRAS.479.2754K}. To make a realistic prediction, the thermal noise is evaluated using realistic configuration of SKA1 Low \citep{SKA1LOW}. We assume a field of view of HSC/PFS to $(15~\rm{deg^2})$\footnote{ The HSC released LAE catalogue at deep/ultra deep fields at $z=6.6$. There are four deep fields and two ultra deep fields. (See {https://hsc.mtk.nao.ac.jp/ssp/survey/}). According to \cite{2021arXiv210402177O}, the LAE catalogue at $z=6.6$ will be available in three deep fields and 2 ultra deep fields. Since one of deep field is at far northern sky (ELAIS-N1 have studied using the LOFAR \citep{2021A&A...648A...7G}.), two deep fields (COSMOS, DEEP2-3) and ultra deep fields (SXDS, COSMOS) would be suitable for the SKA1 Low survey. Thus, in practice, only deep (5.31~$\rm deg^2$, 5.76~$\rm deg^2$) and ultra-deep (2.05~$\rm deg^2$, 2.02~$\rm deg^2$) are currently available. Therefore, we assume a field of view of HSC/PFS to $15~\rm{deg^2}$.}
We first assume that the foregrounds are perfectly removed, and then the error is calculated using all k modes. Next, we use k modes within the EoR window in the assumption of that the foreground contamination is well limited to below the horizon limit.

\begin{figure}

        \includegraphics[width=1\hsize]{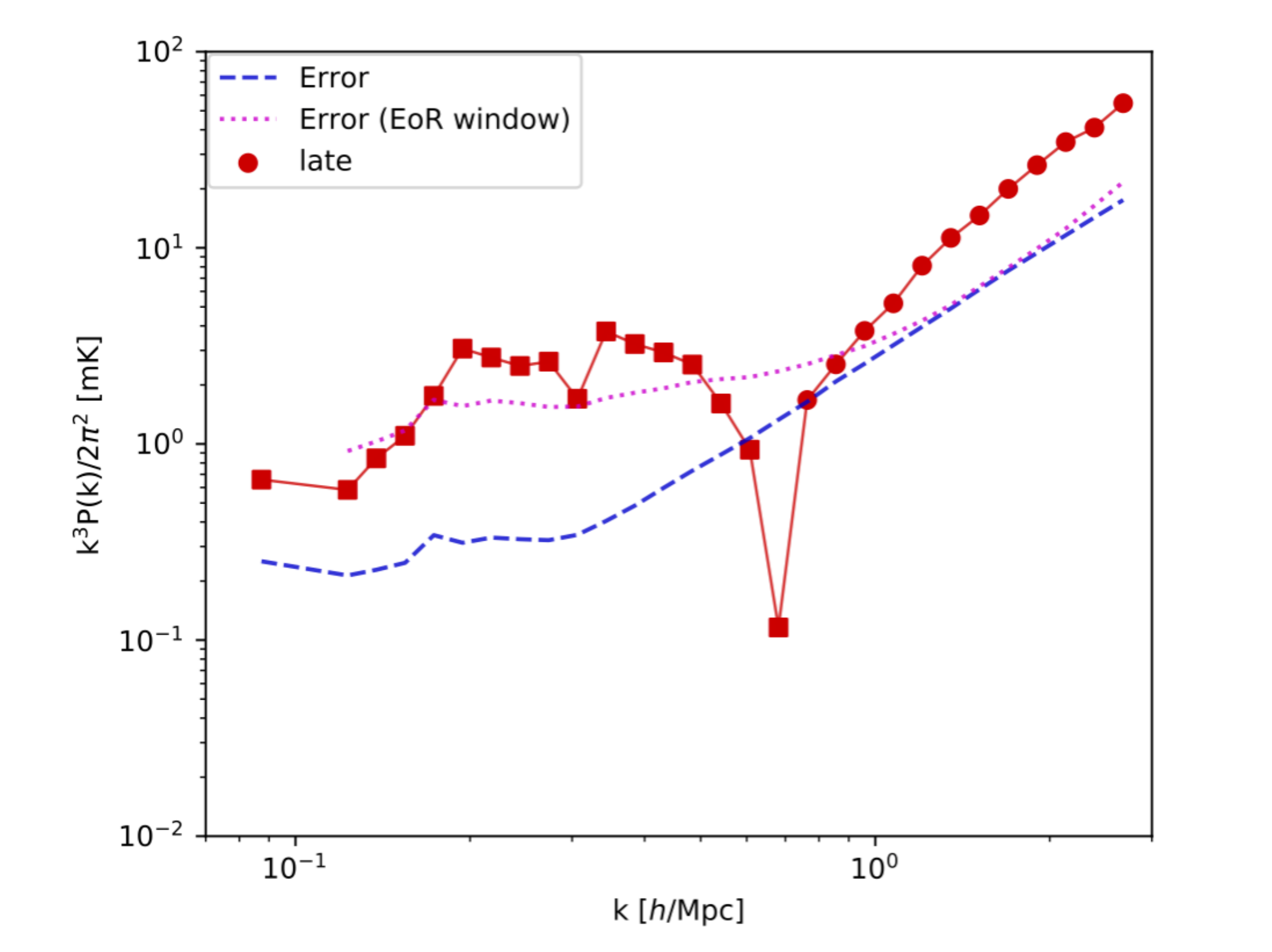}
    \caption{Cross correlation of SKA1 Low and HSC Deep field at $z=6.6$. The SKA1 Low consists of 512 stations. We assume the effective area of antenna $A_e=462\rm{m}^2$ at 150MHz, a bandwidth $B=8\rm{MHz}$, frequency channel width $\Delta \nu =80\rm{kHz}$. Solid line is the absolute value of CPS, and squares and circles are indicate negative and positive value. Dotted line is the error used k modes within the EoR window. The dashed line is also the error where all k-modes are used to evaluate by assuming the perfect foreground removal. Here, we assume 999 hours of the SKA1 Low observation and 15 $\rm deg^2$ of LAE survey area.}
    \label{fig:cps}
\end{figure}

Figure \ref{fig:cps} shows the 21cm-LAE CPS and the errors. Under the assumption of perfect foreground removal, the CPS can be observed at large scales. On the other hand, when adopting the EoR window method, the SNR of CPS becomes worse since the error increases due to the reduction of the number of available k-mode \footnote{
The sensitivity of small scale does not depend on the foreground avoidance, but the signal highly depends on the model of reionization and LAEs (\cite{Kubota2020MNRAS.494.3131K}).}. We also find that only 100 hours of observation per pointings are required to detect the signal once the foreground contamination is correctly removed. On the other hand, 999 hours of observation for each HSC field might be required without foreground removal. As the SKA1 Low has significant high sensitivity, the CPS error can be the cosmic variance dominant at large scales. Thus, a shallow and wide survey might be suitable rather than a deep observation.



\section{Observations of the 21cm line}

Before the summary of this review, we briefly report the current status of the 21cm line observation and future prospects in the SKA1 era. For readers who are interested in current status and data analysis of 21cm observations, we recommend following references to the readers \citep[e.g.][]{2019cosm.book.....M,2020PASP..132f2001L,2021arXiv211006173B}. On the basis of these references, we review current updates of the 21cm observation and future propects in this section. 

\subsection{Foreground and Systematics}
The redshifted 21cm line observed at less than $\nu=$ 200 MHz allows us to explore the dark ages to cosmic reionization. There are a number of radio instruments measuring such low frequencies toward the detection of the 21cm line. For example, the 21cm line global signal which was described in section \ref{subSec:globalsignal} is measurable with a single dipole antenna with long integration time and accurate foreground removal method (see e.g. \cite{2006MNRAS.371..867F,PhysRevD.87.043002}). The global signal observation has been operated by such as the EDGES \citep{EDGESHIGH2017} and SARAS2 \citep{Singh2018SARASReionization2}. We must mention that the EDGES low band analysis showed a strong absorption at $z=17.8$ which was reported in \cite{2018Natur.555...67B}. The result has significant impacts on this field as it has extraordinary feature. At the same time that a number of astrophysical models and cosmologies are suggested to explain the absorption \citep[e.g.][]{Barkana:2018lgd}. Some concerns and systematic errors also have been pointed out; \citet{Hills2018ConcernsData2} early pointed out an unrecognized sinusoidal systematic; \citet{2021MNRAS.502.4405B}, using Maximally Smooth Functions, argued the possibility that  sinusoidal systematic exists in the EDGES low data; \citet{2019ApJ...880...26S} showed that an absorption, consistent standard cosmology, is favored for the EDGES data; \citet{2020MNRAS.492...22S} showed no models of signal and systematics has statistical significance; False signal can be produced from unmodeled beam chromaticity and Galactic diffuse emission \citep{2020ApJ...897..132T}; The EDGES unexpected signal can be explained by ground plane resonances \citep{2019ApJ...874..153B}. Another experiments have operated to detect the 21cm global signal. Most importantly, the best fit absorption obtained in \cite{2018Natur.555...67B} has been rejected by SARAS3 \citep{2022NatAs.tmp...47S}. As the EDGES observation might be suffered from the Radio frequency interference (RFI) and Earth's ionosphere effect, some projects have planed with intent to measure the global signal from the far side of the moon where we can ideally avoid the RFI and the ionospheric effect (e.g. DAPPER; \cite{2021arXiv210305085B}). Such space projects also enable us to measure the 21cm line below 10 MHz, which is not detectable from the ground due to the ionosphere reflection. 

As observation of the 21cm line fluctuation (e.g. the power spectrum described in section~\ref{sec:moment}) needs spatial resolution, radio interferometers have operated such as PAPER \citep{Parsons2010TheResults}, GMRT \citep{Paciga2013AExperiment2}, MWA \citep{Tingay2013PASA,Wayth2018PASA}, LOFAR \citep{VanHaarlem2013LOFAR:Array} and HERA \citep{2017PASP..129d5001D}. The groups of these instruments have published upper limits of the power spectrum at the EoR and the Cosmic Dawn. The detection has not yet been achieved. This is due to the strong foreground and complicated systematics. We will briefly describe these problems below. 

Foreground contamination is dominated by the synchrotron radiation from our Galaxy and extra-galactic radio sources such as AGNs \citep{2008MNRAS.389.1319J,ChapmanJelic2019}. Free-free emission is also a subdominant source of foregrounds. Such emissions are extremely brighter than the expected 21cm signal especially at the direction of the Galactic center and near bright radio sources. Therefore the observation field is usually chosen to be far from the Galactic centre and free of too bright objects. However, even in the fainter fields, foregrounds are more than 3 orders of magnitude brighter than the 21cm signal. Even more, due to the large primary beam and wide side lobes of the low frequency radio telescopes can easily leak the bright sources contamination at far field into the measurement \citep{2016ApJ...819....8P}. Calibration error due to unmodeled sources significantly biases the power spectrum analysis \citep{Barry2016CalibrationDesigned2}. Therefore, understanding and modelling of each foreground source are key challenges. The radio source catalogues in the southern sky, where the SKA1 Low will observe, are available such as GLEAM survey \citep{2015PASA...32...25W,2017MNRAS.464.1146H}. Such a large radio survey is essential as the beam has a response to a large area. The radio catalogue with deep integration toward the targeted field has also been made so far \citep{2017PASA...34...33P,2021arXiv211008400L}. Furthermore, the modelling of extended sources is important as the residuals of extremely bright sources has to be minimized \citep{Line2020ModellingStudy2,2021MNRAS.tmp.2671R}. The radio catalogues are used for either the removal of foreground and instrumental calibration. On the other hand, in addition to the individual extra-galactic source, the diffuse Galactic foreground has to be modeled. The diffuse emission dominates the visibility observed by short baselines. So, such a model will play an essential role to calibrate the short baselines and to correct phased antenna array response. For example, PyGSM \citep{2016ascl.soft03013P} is available to predict the diffuse emission based on previously observed all sky maps. Recently, in \cite{2021arXiv210711487B}, the polarized diffuse emission map was created using MWA data. Such diffuse maps have a great potential to improve the accuracy of calibration.


Rather than avoiding the foregrounds, the removal is essential to detect the signal at large scales. One of the ultimate goals of 21cm line observation is the direct imaging of the 21cm signal in the SKA era, and then the foreground removal will be required with high precision. There are various statistical methods such as polynomial fitting, principal component analysis, FastICA \citep{Chapman2012MNRAS}, GMCA \citep{Chapman2013MNRAS} and GPR \citep{Mertens2018MNRAS;GPR}. The GPR uses knowledge of frequency covariance of the 21cm line, foregrounds, noise and other components to extract the foreground contamination from observed data. These methods should work pretty well based on simulation results. In \cite{2017ApJ...838...65P} and \cite{Mertens2020ImprovedLOFARb}, GMCA and GPR have been applied to the real LOFAR data analysis and showed great improvements to reduce the upper limits of the power spectrum.


The EoR window and the foreground removal can be disturbed by instrumental systematics and calibration error. Thus, the understanding of possible systematics and developments of calibration methods are essential. We describe known systematcs below.

Radio frequency interference (RFI) is a primary source of systematics in any radio observation. To avoid the RFI pollution, the SKA1 Low will be constructed at the MRO which is one of the radio quietest regions. The RFI environment in the MRO was reported in \cite{2015PASA...32....8O}. Although the report shows only a few percent of RFI pollution, careful flagging of RFI is still crucial. The AOFLAGGER has been used in, such as, MWA and LOFAR teams as a powerful tool to identify the RFI \citep{2012A&A...539A..95O}. Furthermore, in \cite{2019PASP..131k4507W2}, they showed that faint RFIs have not been flagged by conventional tools and developed SSINS software. The SSINS identifies the fainter RFIs by integrating the visibility of baselines and comparing the statistical property. Such faint RFIs (e.g. DTV) can bias the power spectrum as shown in \cite{2020MNRAS.498..265W,2021arXiv211008167W}. In \cite{Barry2019ImprovingObservations2}, data flagging using the SSINS shows improvements on the power spectrum upper limits. After the RFI flagging, the treatment of flagged visibility should be taken carefully. For example, in \cite{2019MNRAS.484.2866O}, they showed that the images can have non-smooth spectrum and the power spectrum is biased due to the fluctuation of the weights in the uv plane.

Earth's ionosphere which is ionized atmospheric region from 50 km to 1000 km altitude is also a primary source of systematics of low frequency radio astronomy \citep[e.g.][]{2017isra.book.....T2}. Incoming radio signal is refracted due to the ionosphere. As the result, the phase of visibility is shifted and the shift corresponds to the shift of  the apparent position of radio sources.  Feature of ionosphere has been investigated by MWA \citep[e.g.][]{2015MNRAS.453.2731L} and LOFAR \citep[e.g.][]{2016RaSc...51..927M}. In terms of foreground removal, the subtraction of bright sources could leave residuals. The shift could also affect adversely on the calibration based on the radio catalogue \citep[e.g.][]{2021MNRAS.505.4775Y}. Furthermore, the ionosphere effect is proportional to $1/\nu^2$ at 1st order and higher order terms become important at $\nu<50\rm MHz$ \citep[e.g.][]{DeGasperin2018TheObservations2}. Currently various methods have proposed such as peeling based correction \citep[e.g.][]{2008ISTSP...2..707M,2009A&A...501.1185I,2020A&A...635A.147A}. For example, in the RTS \citep{2008ISTSP...2..707M}, the ionospheric phase is well corrected by comparing the phase of visibility for bright sources with the catalogue based model \citep{2021PASA...38...28C}. Furthermore, influence on the power spectrum might be avoidable by removing the data with active ionosphere \citep{Jordan2017CharacterizationArray2,Trott_2018}.

Accurate modelling of the primary beam response is a crucial challenge for the 21cm observation. The error of the beam model easily affects the direction independent calibration and the foreground removal. The tile of the MWA consists of 16 antennas and the modeling of beam for such phased array is recently developed \citep[e.g.][]{2015RaSc...50...52S}. The model has to include the mutual coupling \citep[e.g.][]{2017PASA...34...62S}. The beam shape would be affected by the status of each involving antenna. Then the beam shape varies with frequency and the variation could produce undesired frequency non-smoothness in the foreground spectrum. Such foreground can contaminate the EoR window and bias the power spectrum measurement \citep[e.g.][]{2020MNRAS.492.2017J}. In particular, the side lobes can leak bright foregrounds far from the pointing center into the visibility. The side lobe can make far side lobe confusion noise and the confusion limits the direct imaging of 21cm line with the SKA1 Low \citep[e.g.][]{2017MNRAS.465.3680M}. Thus, measurement of the phased array is an active research field. For example, measurement of the beam shape has been operated for the MWA \citep[e.g.][]{2015RaSc...50..614N,2018PASA...35...45L,2021MNRAS.502.1990C} and SKA1 Low \citep[e.g.][]{2018ExA....45....1D} using satellites and drones. Antenna Holography proposed in \citep[e.g.][]{2021RaSc...5607171K} could be useful to calibrate the beam model of each SKA1 Low station.

As described, the detection of the 21cm line is suffered from the wide-field beam and the ionospheric refraction. These systematics need to be corrected by performing the direction dependent calibration (DDcal). The DDcal, for example, solves the least-squares problem to determine Jones matrix for each direction of calibrators within the beam. Several teams have developed calibration software such as FHD \citep{2012ApJ...759...17S,2019PASA...36...26B}, RTS\citep{2008ISTSP...2..707M}, OMNICAL \citep{2018ApJ...863..170L}, SAGECAL-CO \citep{2016arXiv160509219Y} and so on. However, the DDcal can leave non-smooth gain structure, which leaks foreground power into the EoR window, if gain is allowed to be variable in frequency. So the gain might be enforced by a polynomial to be smooth \citep[e.g.][]{Mertens2020ImprovedLOFARb}. Furthermore, the DDcal could cause signal loss due to the mismodelling the Galactic diffuse emission \citep[e.g.][]{2016MNRAS.463.4317P}. The signal loss is avoidable by excluding the short baselines from the calibration or creating appropriate models of the diffuse emission \citep{2021arXiv210711487B}.

In addition, for example, the reflection due to impedance mismatch between receiver and cable makes coherent delay signals \citep[e.g.][]{2021MNRAS.500.1232F}. Beam calibration error can leak polarized emission into stokes I \citep[e.g.][]{2015MNRAS.451.3709A}. The leaked emission, even worth, can be non-smooth in frequency due to the Faraday rotation. We emphasize that the possible systematics are not limited to things described above. As the upper limit of the power spectrum becomes smaller, more unknown systematics may be found which prevent the detection of the power spectrum.

\subsection{Current results}

\begin{figure*}
    \includegraphics[width=170mm]{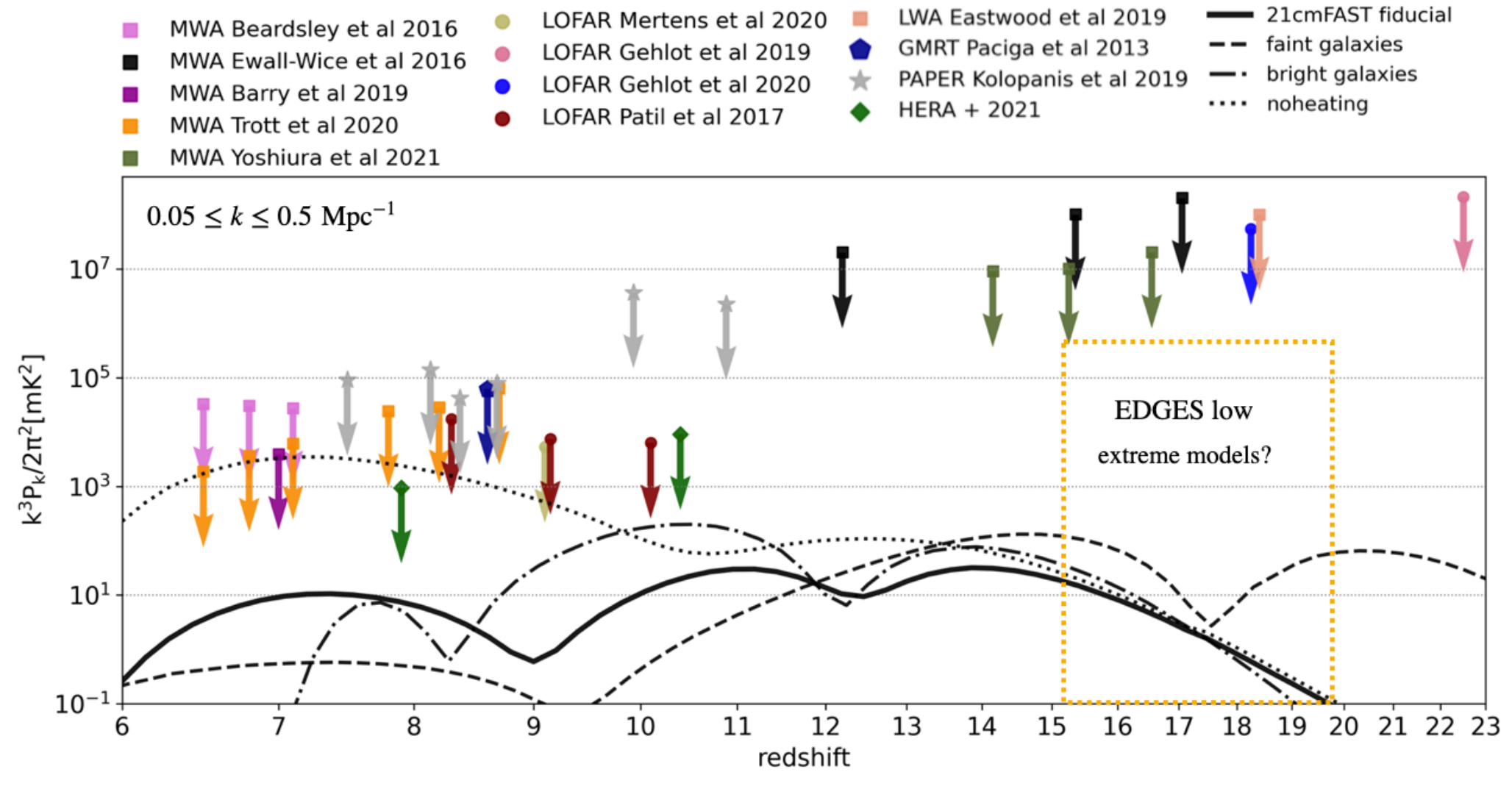}
    \caption{Current upper limits on the 21cm line power spectrum at a range of redshifts. We also show the power spectrum at $k=0.1~{\rm Mpc^{-1}}$ as references. Solid line is a fiducial model of 21cmFAST. Dashed and dot-dashed lines are the faint galaxy model and the bright galaxy model given in \cite{2016MNRAS.459.2342M}. Dotted line is the no-heating model which conflicts with the HERA's result. The extremely strong absorption was reported at $z=17.8$ in \cite{2018Natur.555...67B}, and the extreme models interpreting the EDGES result predict strong fluctuations \citep[e.g.][]{Fialkov2019SignatureSpectrum2}. This figure is made by referencing figure.~7 in \cite{2020PASP..132f2001L} and figure.~2 in \cite{2021arXiv211006173B}.}
    \label{fig:eorupper}
\end{figure*}

Using the calibrated and foreground removed visibility, the power spectrum has been evaluated with power spectrum estimation tools (e.g. CHIPS;\cite{Trott2016CHIPS:ESTIMATOR2}). While the detection of the 21cm power spectrum has not been achieved, the upper limits on the power spectrum have been reported at various redshifts and scales: PAPER\citep{2019ApJ...883..133K}, GMRT\citep{Paciga2013AExperiment2}, LWA\citep{2019AJ....158...84E,2021MNRAS.506.5802G}, MWA\citep{2015PhRvD..91l3011D,2016MNRAS.460.4320E,2016ApJ...833..102B,2019ApJ...887..141L,Barry2019ImprovingObservations2,Trott2020DeepObservations,2021MNRAS.504.2062P,2021MNRAS.505.4775Y,2021MNRAS.tmp.2671R}, LOFAR\citep{Patil2017,2019MNRAS.488.4271G,Mertens2020ImprovedLOFARb,2020MNRAS.499.4158G} and HERA\citep{2021arXiv210802263T}. Figure ~\ref{fig:eorupper} compares the upper limits and theoretical models of power spectrum at a range of redshifts. For the reference, we show the evolution of the power spectrum of fidutial model of 21cmFAST, faint galaxy model, bright galaxy model \citep{2016MNRAS.459.2342M} and no-heating model (i.e. spin temperature is coupled with gas temperature which cools adiabatically). Current upper limits are still more than 2 orders of magnitude larger than the fiducial models. However, the upper limits at the EoR have started to constrain the astrophysical models \citep{Greig2021MNRASMWA,2021MNRAS.503.4551G,Greig2021MNRASLOFAR,2020MNRAS.498.4178M,2021arXiv210807282T}. The results indicate no-heating model of reionization is strongly disfavored. Thus, for example, to moderately heat up the IGM, the X-ray luminosity of the source needs to be higher than that of the local source \citep{2021arXiv210807282T}. This is partly demonstrated in the Figure.~\ref{fig:eorupper}. The no-heating model is inconsistent with the HERA's upper limit. 

At the cosmic dawn ($12<z<23$), the most tight upper limits were given in \cite{2021MNRAS.505.4775Y}, where they processed more than 10 hours of MWA data at $75\le \nu \le 100~\rm MHz$ and at two MWA EoR fields which are far from the Galactic center. They found that low quality DDcal leaves non-smooth foreground residuals after the source subtraction, and therefore they turned off the DDcal in the data reduction. Furthermore, as the data suffered from strong ionospheric refraction and RFI contamination at the FM band, careful data selection was performed and roughly 3 hours of clean data sets were found for each field. The resultant upper limits, however, are still 5 orders of magnitude larger than the standard 21cm signal and systematic dominant. Even if looking at other projects, the upper limits are also not tight enough to constrain any astrophysical models. This could be because (i) understanding of systematics is not well at the frequency because previous efforts are focused on the EoR frequency (ii) foregrounds becomes powerful at the cosmic dawn frequency due to the spectral index of synchrotron emission (iii) ionospheric phase shift is proportional to $\nu^{-2}$ and (iv) significant thermal noise disturbs the calibration. It would be worth to mentioning that some theoretical models explaining the strong absorption of the EDGES result predict significant enhancement of the power spectrum \citep[e.g.][]{Fialkov2019SignatureSpectrum2}. Thus, the data analysis of the cosmic dawn frequency has recently become more attractive. 

While the power spectrum analysis has been actively conducted, the bispectrum was also evaluated using MWA data and obtained thermal noise limited results for some triangles \citep{2019PASA...36...23T}. Furthermore other statistical methods have been attempted using actual radio data. For example, using MWA data and Subaru LAE catalogue, the pre-whitening matched filter detector was used to constrain the IGM brightness temperature \citep{2021MNRAS.507..772T}. Using the HERA data, the bispectrum phase has been tested to detect the 21cm signal avoiding systematics \citep{2020PhRvD.102b2002T}.


\subsection{Future Prospects}
The theoretical sensitivity for the observation of the power spectrum, bispectrum and cross power spectrum has been estimated in for example \cite{McQuinn:2005hk},\cite{2015MNRAS.451..266Y}, \cite{Kubota2018MNRAS.479.2754K} and \cite{Weinberger2020}. In terms of foreground avoidance, because some k-modes are not available, the scales to be observed are limited. Nevertheless, the ongoing instruments such as MWA, LOFAR and HERA have enough sensitivity to detect the 21cm power spectrum. As the power spectrum analysis is the primary stream of 21cm line analysis, the astrophysical model would be first tested in terms of the power spectrum. However, the upper limits of the 21cm line power spectrum is limited to systematics rather than thermal noise (although the recent HERA result is consistent with the noise). Therefore, even for the SKA1 Low, further developments on the calibration are required. 

Because the 21cm line observation is severely contaminated with astrophysical foregrounds and instrumental systematics, the validation of future detected signals would also be a huge challenge. For example cross pipeline detection would be required to avoid the systematics caused by analysis \citep[e.g. ][]{2016ApJ...825..114J}. Thus, the development of multiple calibration software, multiple foreground removal methods and multiple statistics is essential and is being actively conducted. Other than that, observation of different sky fields would be required to validate the bias due to the foregrounds contamination. Observation with the precursors of the SKA1 should be proceeded as the detection of the signal with different instruments is also important to avoid the fake detection due to the instrumental systematics. Furthermore, as mentioned in Sec.~\ref{sec:21cmCC}, the cross correlation with external objects would definitely be strong evidence of the 21cm line signal.

As we show in Figure.~\ref{fig:eorupper}, the upper limits on the power spectrum are getting reduced. The improvements have been achieved thanks to longer integration, massive instruments and most importantly the understanding of the systematics and active development of software. Further upgrade of current instruments has been planned and done (e.g. LOFAR 2.0, MWA Phase III and HERA full operation). Thus, we could expect the upper limits will be improved in the next 5 years. Furthermore, construction of SKA1 Low has begun in 2021, and the full operation is scheduled to begin in 2028 $\sim$ 2029\footnote{https://www.skatelescope.org/news/green-light-for-ska-construction}. Therefore, the observation of the 21cm line by the SKA1 Low would bring significant scientific results in the 2030s.

\section{Summary}
The epoch from the dark ages to cosmic reionization is the frontier of the universe. Current galaxy observations conducted by powerful telescopes such as subaru and ALMA are revealing the end stage of the EoR. However, we need other approaches to explore more higher redshift beyond the end of the EoR. The cosmic 21cm line signal is a powerful tool to explore the epoch. To exploit fruitful information from cosmological 21cm line signal, it is necessary to understand the properties of the 21cm line signal and to establish the method to extract useful information from the 21cm line signal. In this paper, we reviewed current progress of cosmic dawn and EoR studies. At first, we introduced basic physics of the 21cm line and then we review the impact of first stars on the 21cm line signal (Chapter 2 $\&$ 3). To extract information from 21cm line signal is crucial to interpret the 21cm line signal. For this purpose, we introduced statistical and machine approaches as methods to extract information from 21cm line signals in chapter 4. Not only 21cm line signal, but also synergy between 21cm line signal and other lines is also important. We introduced the cross-correlation between 21cm line signal and other wavelength observations as a method to consider the synergy between other emission lines and the 21cm line signal in chapter 5. Finally, we reviewed the aspects of the 21cm line experiments in chapter 6. We summarized observational challenges such as foreground and systematics and reported current constraints on the 21cm line power spectrum.

\section*{Acknowledgement}

HS is supported by National SKA Program of China (No. 2020SKA0110401) and National Natural Science Foundation of China (No.12103044). 
This work was supported in part by JSPS KAKENHI Grant Nos. 21J00416(SY), 18K03699(KH), 17H04827, 20H04724, 21H04489 (HY),
NAOJ ALMA Scientific Research Grant Numbers 2019-11A (HY),
 JST FOREST Program JPMJFR20352935 (KH) and JPMJFR202Z (HY). SY is supported by JSPS Research Fellowships for Young Scientists.

\bibliographystyle{apj}
\bibliography{reference,eorSY,example}

\end{document}